\documentclass[twocolumn]{aa}  
\usepackage{natbib}
\usepackage{subcaption}
\bibpunct{(}{)}{;}{a}{}{,}
\usepackage{booktabs,caption}

\newcommand\gaia{\textit{Gaia}}

\defcitealias{massari23}{Paper I}
\defcitealias{niederhofer25}{Paper IV}
\defcitealias{aguado25}{Paper II}
\defcitealias{ceccarelli25}{Paper III}
\defcitealias{zerbinati26}{Paper VI}

\title{Cluster Ages to Reconstruct the Milky Way Assembly (CARMA)}
\subtitle{V. The chronological merger tree of the Milky Way}

\titlerunning{CARMA V.}
\authorrunning{Aguado-Agelet et al.}
\author{Fernando Aguado-Agelet \inst{1,2}, 
Chiara Zerbinati \inst{3,4},
Davide Massari \inst{3}, 
Matteo Monelli \inst{5,6,2}, 
Santi Cassisi \inst{5,7}, 
Cristiano Fanelli \inst{3},
Carme Gallart \inst{6,2}, 
Edoardo Ceccarelli \inst{8,3,4},
Yllari Kay González Koda \inst{9},
Tom\'as Ruiz-Lara \inst{9,11}, 
Sara Saracino \inst{10}, 
Maurizio Salaris \inst{3,10}
}
\institute{              atlanTTic, Universidade de Vigo, Escola de Enxeñar\'ia de Telecomunicaci\'on, 36310, Vigo, Spain\\ \email{faguado@uvigo.gal}
             \and
             Universidad de La Laguna, Avda. Astrof\'isico Fco. S\'anchez, E-38205 La Laguna, Tenerife, Spain
             \and
             INAF - Osservatorio di Astrofisica e Scienza dello Spazio di Bologna, Via Gobetti 93/3, I-40129 Bologna, Italy
             \and
             Dipartimento di Fisica e Astronomia, Universit\`a degli Studi di Bologna, Via Piero Gobetti 93/2, 40129 Bologna, Italy\\ \email{chiara.zerbinati2@unibo.it}
             \and
             INAF – Osservatorio Astronomico di Abruzzo, Via M. Maggini, 64100 Teramo, Italy
             \and
             Instituto de Astrof\'isica de Canarias, Calle V\'ia L\'actea s/n, E-38206 La Laguna, Tenerife, Spain
             \and
             INFN - Sezione di Pisa, Universit\'a di Pisa, Largo Pontecorvo 3, 56127 Pisa, Italy
             \and
             Kapteyn Astronomical Institute, University of Groningen, P.O. Box 800, 9700 AV Groningen, The Netherlands
             \and
             Universidad de Granada, Departamento de Física Teórica y del Cosmos, Campus Fuente Nueva, Edificio Mecenas, 18071 Granada, Spain
             \and
             Astrophysics Research Institute, Liverpool John Moores University, 146 Brownlow Hill, Liverpool L3 5RF, UK
             \and
             Instituto Carlos I de F\'isica Te\'orica y Computacional, Facultad de Ciencias, E-18071 Granada, Spain
}

\abstract{We present a new age determination of 24 globular clusters (GCs) dynamically associated with the main accretion events experienced by the Milky Way (MW), as part of the Cluster Ages to Reconstruct the Milky Way Assembly (CARMA) project's effort to trace the Galaxy's assembly history. We used deep and homogeneous archival {\it Hubble} Space Telescope data, and applied the CARMA isochrone-fitting code to derive homogeneous estimates of age, metallicity, reddening, and distance modulus for systems dynamically associated with Gaia-Sausage-Enceladus (GSE), the Sagittarius dwarf galaxy (Sag), the Helmi streams (H99), and the Sequoia galaxy (Seq). These 24 new determinations are supplemented by 11 previously studied GSE clusters to construct the complete age-metallicity relation (AMR) of the GSE system.
We find that each progenitor system describes a well-defined AMR, with a distinct slope and extent reflecting its individual star-formation efficiency and chemical enrichment history. By fitting analytical AMR models within a Markov chain Monte Carlo framework, we quantify the stellar mass and accretion time for each progenitor galaxy. This results in the first detailed merger tree obtained from strictly homogeneous chronological information, according to which the Low-energy-Kraken-Heracles (LKH) system is the first merger experienced by the MW that brought GCs in, followed by Sequoia, H99, GSE, and finally Sgr. The most significant events in terms of stellar mass are LKH, GSE, and Sgr, which together contribute a total of $\sim2.5\times10^9\,\mathrm{M}_\odot$. This corresponds to more than 95\% of the stellar mass accreted by the MW from mergers massive enough to host GCs.}


\providecommand\gaia{\textit{Gaia}}

\begin{document}
\flushbottom
\maketitle
\thispagestyle{empty}

\section{Introduction}

The quest to reconstruct the Milky Way (MW) assembly history is experiencing a revolution, driven by the unprecedented phase-space information provided by the {\it Gaia} space mission \citep{gaia16, gaiadr2, gaiadr3}. In combination with the line-of-sight velocities and chemical abundances from large-scale spectroscopic surveys, this wealth of kinematic data has allowed important advances in the discovery and characterisation of the past merger events that have shaped our Galaxy into its current appearance \citep[see][for a review]{helmi20}. It is now widely accepted that the local halo is dominated by the debris of the last major merger experienced by the MW, involving the Gaia-Sausage-Enceladus (GSE) dwarf galaxy \citep{helmi18, belokurov18}. Alongside GSE, numerous other accreted substructures have been identified, including the remnants of the Sagittarius dwarf galaxy \citep{ibata94}, the Sequoia galaxy \citep{myeong19}, the Helmi streams \citep{helmi99}, and various other coherent structures in the integrals of motion space \citep[e.g.][]{massari19, koppelman19, naidu20, horta21, callingham22, malhan22, oria22, tenachi22, belokurov22, mikkola23, dodd23, massari26}.

However, tracing the origin of stellar populations based purely on dynamical properties is often challenging, as the debris of different merger events may overlap, and coherent dynamical substructures can be neither pure nor complete \citep{buder22, rey23, chen24, mori24, thomas25, akib26}. To break this ambiguity, dynamical selections must be complemented by additional conserved quantities, such as age and chemical abundances. In this context, globular clusters (GCs) stand out as invaluable probes. As the oldest observable stellar systems, GCs formed in accreted dwarf galaxies exhibit distinct age-metallicity relations (AMRs) that differentiate them from those followed by in situ populations \citep[e.g.][]{marinfranch09, forbes10, leaman13, massari19}. Consequently, the precise age determination of GCs provides a robust method of disentangling the complex accretion history of the MW and characterising the physical properties of its building blocks, such as their mass and star-formation history (SFH; \citealt{kruijssen19}).

Recognising the critical role of ages in Galactic archaeology, we established the Cluster Ages to Reconstruct the Milky Way Assembly (CARMA) framework. CARMA aims to determine precise relative ages for the entire system of known MW GCs, defining a systematic-free chronological scale by employing homogeneous deep photometry, state-of-the-art theoretical isochrones, and robust statistical methods. In the previous papers of this series, we successfully applied the CARMA framework to resolve the debated origin of metal-rich bulge GCs \citep[][Paper I]{massari23}, constructed the AMR of the GSE GC system \citep[][Paper II]{aguado25}, identified the first Splashed cluster, NGC~288 \citep[][Paper III]{ceccarelli25}, studied the chrono-dynamical structure of old clusters in the Large Magellanic Cloud (LMC), finding robust evidence for the accretion of NGC~1841 from a smaller satellite \citep[][Paper IV]{niederhofer25}, and derived the AMR of the in situ GC population to constrain the age of the Universe \citep[][Paper VI]{zerbinati26}. In addition, \citet{massari26} provided the strongest evidence for the existence of the earliest massive merger known to date, whose debris is currently buried in the inner Galaxy, and named Low-energy-Kraken-Heracles \citep[LKH; see also][]{massari19, kruijssen20, horta21}. These studies have demonstrated the power of the CARMA methodology in confirming the origin of individual GCs at any metallicity and reconstructing the properties of their progenitor galaxies. The progenitor assignments and the CARMA age estimates are continuously updated as new data become available, and the most recent version of the full catalogue is publicly accessible at the CARMA web page\footnote{\url{https://www.oas.inaf.it/en/research/m2-en/carma-en/}} \citep{massari25}.

Beyond GSE, other prominent accreted systems, namely Sagittarius, the Helmi streams, and Sequoia, are considerably less well characterised in terms of their individual SFHs, stellar masses, and accretion timescales. The Sagittarius (Sgr) dwarf spheroidal is the most prominent ongoing merger event, with its tidal streams encircling the entire sky \citep{ibata94, majewski03}. Its progenitor had a total initial stellar mass of $\sim10^8$--$10^9$\,M$_\odot$ \citep{kruijssen19, vasiliev21sgr}, and its SFH reveals a complex, extended sequence in age-metallicity space, with the star-formation rate (SFR) dropping sharply approximately 5--7 Gyr ago, possibly triggered by its interaction with the MW \citep{siegel2007, deboer15, ruizlara20}. The Helmi streams \citep[H99;][]{helmi99} represent the debris of a dwarf galaxy with a stellar mass of $\sim10^8$\,M$_\odot$, accreted between 8 and 11 Gyr ago \citep{koppelman19, ruizlara22}. Seven GCs have been associated with this progenitor \citep{massari25}, of which four have deep archival \textit{Hubble} Space Telescope (HST) photometry available and are studied in this work (NGC~4590, NGC~5053, NGC~6426, and Rup~106). Finally, the Sequoia galaxy \citep{myeong19} contributed the bulk of the high-energy retrograde stars in the Galactic halo, with an estimated stellar mass of $\sim5\times10^8$\,M$_\odot$ and a total mass of $\sim10^{10}$\,M$_\odot$. Its stars are characterised by lower metallicities and lower $\alpha$-element abundances compared to GSE at similar energies \citep{matsuno22, ceccarelli24a}, suggesting a lower SFR and hence slower chemical enrichment \citep{myeong19, dodd25}. Despite their significance, the precise AMRs for these three systems remain largely unconstrained in a homogeneous framework, hindering a direct comparison of their evolutionary properties with those of GSE.

In this fifth paper of the series, we extend the application of our isochrone-fitting framework to a comprehensive sample of 24 GCs dynamically associated with the main accretion events experienced by the MW, excluding the LMC. While previous works focused specifically on the GSE system, this study broadens the scope to include clusters linked to Sagittarius, the Helmi streams, and Sequoia, as well as refining the associations of several GSE candidates. By deriving precise AMRs for each of these progenitor systems, we aim to constrain their individual SFHs and accretion timescales, ultimately placing them within a coherent chronological merger tree of the MW.

The paper is organised as follows. Section~\ref{sec:sample} describes the selected GC sample and methods. Section~\ref{sec:amr_modelling} presents the analytical framework for AMR modelling. Section~\ref{sec:results} presents the results, including the membership assignments, the characterisation of each progenitor, and comparisons with field star ages and the LMC. Section~\ref{sec:merger_tree} presents the chronological merger tree of the MW. A summary and conclusions are provided in Sect.~\ref{sec:summary}.

\section{GC sample and methods}\label{sec:sample}

The approach adopted in this work strictly follows the analysis presented in \citetalias{massari23} and \citetalias{aguado25}, and extends it to a comprehensive sample of accreted GCs associated with several distinct progenitor dwarf galaxies (specifically GSE, Sagittarius, Helmi streams, Sequoia, and Elqui). One of the most important pillars of the CARMA project is homogeneity, both in the data (source, data reduction, calibration) and in the theoretical models and statistical methods employed. We briefly summarise the specific aspects relevant for the current work.

\begin{table*}[!htbp]
\centering
\caption{\label{tab:results} Results of the isochrone fitting for the 24 GCs in our sample.}

\begin{tabular}{lccccc}
Name & [M/H] & E(B-V) & DM & Age & Progenitor  \\
& & [mag] & [mag] & [Gyr] &  \\
\hline
\\
\hline \\[-6pt]
\multicolumn{6}{l}{Gaia-Sausage-Enceladus (GSE)} \\[2pt]
\hline \\[2pt]
NGC~4147 & -1.52 \(^{+0.05}_{-0.02}\) & 0.00 \(^{+0.01}_{-0.01}\) & 16.43 \(^{+0.01}_{-0.01}\) & 13.06 \(^{+0.13}_{-0.30}\) & GSE \\[3pt]
NGC~5024 & -1.76 \(^{+0.03}_{-0.01}\) & 0.02 \(^{+0.01}_{-0.01}\) & 16.25 \(^{+0.01}_{-0.01}\) & 13.85 \(^{+0.10}_{-0.10}\) & GSE \\[3pt]
NGC~5272 & -1.24 \(^{+0.04}_{-0.19}\) & 0.01 \(^{+0.01}_{-0.01}\) & 14.97 \(^{+0.02}_{-0.02}\) & 12.48 \(^{+0.48}_{-0.41}\) & GSE \\[3pt]
NGC~5904 & -1.06 \(^{+0.02}_{-0.07}\) & 0.03 \(^{+0.01}_{-0.01}\) & 14.34 \(^{+0.01}_{-0.01}\) & 11.82 \(^{+0.32}_{-0.14}\) & GSE \\[3pt]
NGC~6101 & -1.69 \(^{+0.05}_{-0.05}\) & 0.10 \(^{+0.01}_{-0.01}\) & 15.74 \(^{+0.01}_{-0.01}\) & 13.75 \(^{+0.13}_{-0.15}\) & GSE \\[3pt]
NGC~6584 & -1.21 \(^{+0.02}_{-0.03}\) & 0.09 \(^{+0.01}_{-0.01}\) & 15.61 \(^{+0.01}_{-0.01}\) & 12.71 \(^{+0.30}_{-0.29}\) & GSE \\[3pt]
NGC~6934 & -1.31 \(^{+0.10}_{-0.08}\) & 0.10 \(^{+0.01}_{-0.01}\) & 15.91 \(^{+0.01}_{-0.01}\) & 13.14 \(^{+0.06}_{-0.07}\) & GSE \\[3pt]
NGC~6981 & -1.24 \(^{+0.04}_{-0.03}\) & 0.04 \(^{+0.01}_{-0.01}\) & 16.12 \(^{+0.01}_{-0.01}\) & 12.70 \(^{+0.29}_{-0.25}\) & GSE \\[3pt]
NGC~7006 & -1.28 \(^{+0.08}_{-0.01}\) & 0.07 \(^{+0.01}_{-0.01}\) & 17.99 \(^{+0.02}_{-0.03}\) & 12.71 \(^{+0.13}_{-0.10}\) & GSE \\[3pt]
Pal~2    & -1.24 \(^{+0.05}_{-0.08}\) & 1.14 \(^{+0.01}_{-0.01}\) & 17.11 \(^{+0.01}_{-0.01}\) & 12.52 \(^{+0.53}_{-0.49}\) & GSE \\[3pt]
Pal~15   & -1.76 \(^{+0.02}_{-0.03}\) & 0.40 \(^{+0.01}_{-0.01}\) & 18.23 \(^{+0.01}_{-0.01}\) & 13.92 \(^{+0.14}_{-0.10}\) & GSE \\[3pt]
\hline \\[-6pt]
\multicolumn{6}{l}{Sagittarius (Sgr)} \\[2pt]
\hline \\[2pt]
Arp~2    & -1.51 \(^{+0.03}_{-0.03}\) & 0.08 \(^{+0.01}_{-0.01}\) & 17.37 \(^{+0.02}_{-0.02}\) & 13.00 \(^{+0.49}_{-0.50}\) & Sgr \\[3pt]
Pal~12   & -0.78 \(^{+0.05}_{-0.05}\) & 0.02 \(^{+0.01}_{-0.01}\) & 16.38 \(^{+0.01}_{-0.01}\) &  9.29 \(^{+0.33}_{-0.30}\) & Sgr \\[3pt]
Terzan~7 & -0.58 \(^{+0.03}_{-0.02}\) & 0.06 \(^{+0.01}_{-0.01}\) & 16.99 \(^{+0.01}_{-0.01}\) &  8.07 \(^{+0.19}_{-0.11}\) & Sgr \\[3pt]
Terzan~8 & -1.78 \(^{+0.10}_{-0.09}\) & 0.10 \(^{+0.01}_{-0.01}\) & 17.18 \(^{+0.01}_{-0.01}\) & 14.16 \(^{+0.26}_{-0.30}\) & Sgr \\[3pt]
\hline \\[-6pt]
\multicolumn{6}{l}{Helmi Streams (H99)} \\[2pt]
\hline \\[2pt]
NGC~4590 & -1.92 \(^{+0.11}_{-0.09}\) & 0.05 \(^{+0.01}_{-0.01}\) & 15.01 \(^{+0.02}_{-0.01}\) & 12.98 \(^{+0.12}_{-0.17}\) & H99 \\[3pt]
NGC~5053 & -1.84 \(^{+0.14}_{-0.17}\) & 0.01 \(^{+0.01}_{-0.01}\) & 16.15 \(^{+0.02}_{-0.01}\) & 13.18 \(^{+0.26}_{-0.28}\) & H99 \\[3pt]
NGC~6426 & -1.82 \(^{+0.03}_{-0.04}\) & 0.37 \(^{+0.01}_{-0.01}\) & 16.55 \(^{+0.01}_{-0.01}\) & 13.12 \(^{+0.30}_{-0.31}\) & H99--GSE \\[3pt]
Rup~106  & -1.41 \(^{+0.12}_{-0.12}\) & 0.17 \(^{+0.01}_{-0.01}\) & 16.60 \(^{+0.01}_{-0.01}\) & 11.64 \(^{+0.49}_{-0.56}\) & H99 \\[3pt]
NGC~7078 & -1.85 \(^{+0.02}_{-0.02}\) & 0.08 \(^{+0.00}_{-0.00}\) & 15.08 \(^{+0.01}_{-0.01}\) & 13.34 \(^{+0.36}_{-0.27}\) & H99--GSE \\[3pt]
\hline \\[-6pt]
\multicolumn{6}{l}{Sequoia (Seq)} \\[2pt]
\hline \\[2pt]
IC~4499  & -1.41 \(^{+0.09}_{-0.03}\) & 0.21 \(^{+0.01}_{-0.01}\) & 16.39 \(^{+0.01}_{-0.01}\) & 12.27 \(^{+0.16}_{-0.18}\) & Seq--GSE \\[3pt]
NGC~5466 & -1.69 \(^{+0.06}_{-0.01}\) & 0.00 \(^{+0.01}_{-0.01}\) & 16.00 \(^{+0.01}_{-0.02}\) & 13.17 \(^{+0.10}_{-0.28}\) & Seq--GSE \\[3pt]
NGC~3201 & -1.36 \(^{+0.05}_{-0.03}\) & 0.26 \(^{+0.01}_{-0.01}\) & 13.34 \(^{+0.01}_{-0.01}\) & 12.36 \(^{+0.33}_{-0.29}\) & Seq--GSE \\[3pt]
\hline \\[-6pt]
\multicolumn{6}{l}{Elqui (Elq)} \\[2pt]
\hline \\[2pt]
Pyxis    & -1.07 \(^{+0.03}_{-0.02}\) & 0.27 \(^{+0.01}_{-0.01}\) & 17.88 \(^{+0.01}_{-0.02}\) & 10.61 \(^{+0.28}_{-0.17}\) & Elq \\[3pt]

\end{tabular}
\tablefoot{DM denotes the true distance modulus. The `Progenitor' column lists the final assignments of this work: the dynamical associations of \citet{massari25} updated following the AMR-based membership analysis of Sect.~\ref{sec:membership}. Double notation (e.g. Seq--GSE) indicates clusters whose association remains uncertain (see Table~\ref{tab:probabilities}).}
\end{table*}

\subsection{Accreted globular cluster sample}
Our sample consists of 24 MW GCs whose dynamical properties associate them with the main accretion events experienced by the Galaxy. The associations are based on the classification by \citet{massari19}, who combined integrals-of-motion analysis with age and metallicity information to link individual GCs to known progenitor systems, including GSE, the Sagittarius dwarf galaxy (Sgr), the Helmi streams (H99), and the Sequoia galaxy (Seq). We adopt here the updated version  by \citet{massari25}, which incorporates the improved proper motions from the \textit{Gaia} early third data release \citep[eDR3;][]{gaiaEDR3}.

The final sample includes all GCs for which deep archival HST photometry is publicly available, either in the HUGS catalogue or in the ACS Survey of Galactic Globular Clusters \citep{sarajedini07}. The only two clusters excluded from the analysis are NGC~5139 ($\omega$~Cen) and NGC~6715 (M54), which will be analysed in a dedicated forthcoming work because of the complexity of their stellar populations. Clusters from the third large public HST survey on GCs (the Missing Globular Cluster Survey, see \citealt{massari26} and Libralato et al., in prep.) will be presented in a forthcoming paper (Dodd et al. in preparation). The full list of clusters, together with their progenitor assignments and the derived isochrone-fitting parameters, is provided in Table~\ref{tab:results}. We stress that the 'Progenitor' column of Table~\ref{tab:results} does not simply reproduce the dynamical classification of \citet{massari25} used to select the sample: it lists our final assignments, in which the ambiguous cases have been resolved (or labelled with double notation) according to the AMR-based analysis presented in Sect.~\ref{sec:membership}. We highlight already at this point that Pyxis is associated with the Elqui (Elq) stream \citep{massari19, massari25}, a high-energy retrograde accreted substructure. As it is the only GC in our sample dynamically linked to this progenitor, no MCMC AMR modelling is performed for this system, and it is therefore excluded from the chronological merger tree presented in Sect.~\ref{sec:merger_tree}; its isochrone-fitting parameters are listed in Table~\ref{tab:results} and briefly discussed in Sect.~\ref{sec:mcmc_results}.

\subsection{Photometry}
The photometric dataset used in this work comes from two complementary HST surveys, both providing imaging with the Advanced Camera for Surveys (ACS) in the F606W and F814W filters. For 12 GCs in our sample, we used the HST UV Globular Cluster Survey \citep[HUGS;][]{piotto15, nardiello18}, adopting the $F606W$ and $F814W$ catalogues labelled as `method-2' \citep[see also][]{anderson08}. For the remaining 12 GCs (predominantly clusters associated with Sagittarius and other systems not covered by the HUGS programme), we used the photometric catalogues from the ACS Survey of Galactic Globular Clusters \citep{sarajedini07}, processed following the same prescriptions as for the HUGS data. As described in detail in \citetalias{massari23} and \citetalias{aguado25}, the F606W$+$F814W photometric system offers two key advantages: it provides the deepest available colour-magnitude diagram (CMD) common to virtually all MW GCs, and the optical bandpass combination is the least sensitive to the presence of multiple evolutionary sequences caused by chemical peculiarities in light elements, in the so-called multiple-population phenomenon \citep[e.g.][]{cassisi20}.

In both cases, the catalogues were processed homogeneously in order to obtain the cleanest possible sample of stars for a reliable comparison with theoretical isochrones: (i) only stars with a membership probability larger than 90\% were retained; (ii) the apparent magnitudes were corrected for differential reddening following the procedure described in \citetalias{ceccarelli25} \citep[based on the method of][]{milone12}; (iii) sources in the innermost regions were removed to avoid poor photometric measurements due to crowding (20 to 60\arcsec, depending on the cluster); (iv) when multi-band photometry was available, highly peculiar populations (such as those enriched in C+N+O or He) were removed by identifying them in the multi-band chromosome maps \citep{milone17}. 

\subsection{The CARMA isochrone-fitting code}
The relative age derivation was performed using the CARMA isochrone-fitting code, originally developed by \citet{saracino19} and described in full detail in \citetalias{massari23}. We briefly summarise the specific aspects relevant for the current work. The code implements a Markov chain Monte Carlo (MCMC) framework to simultaneously determine the best-fitting age, global metallicity [M/H], colour excess E(B$-$V), and true distance modulus (DM), by performing a star-by-star comparison of the observed CMD with a grid of theoretical isochrones.

We adopted models from the latest release of the BaSTI stellar evolution library \citep{hidalgo18, pietrinferni21}, which cover a fine grid in age (6 to 15 Gyr, in steps of 100 Myr) and metallicity ($-2.5$ to $+0.3$ dex, in steps of 0.01 dex). Solar-scaled models including diffusion were used throughout. The adoption of solar-scaled models avoids making assumptions on the $\alpha$-element abundance, which is often poorly constrained or systematically uncertain. Our choice is justified by the fact that in these optical bands, solar-scaled and $\alpha$-enhanced isochrones at a fixed global metallicity [M/H] and age are effectively equivalent, as demonstrated in \citet{cassisi04}. By working in global metallicity [M/H] rather than [Fe/H], we absorb the $\alpha$-element contribution through the well-established relation \citep{salaris93, cassisi04}
\begin{equation}
    [M/H] = [Fe/H]+\log(0.694\times10^{[\alpha/Fe]}+0.306).
\end{equation}
For clusters with E(B$-$V) $>$ 0.1 mag, we additionally applied temperature-dependent reddening corrections to the isochrones \citep{girardi08}.

Gaussian priors on [M/H] and E(B$-$V) are centred on the values from the \citet{harris10} catalogue (unless recent high-resolution spectroscopic abundances became available, as discussed later), while the prior on DM is centred on the determinations from \citet{baumgardt21}. Their standard deviations are set individually from the literature uncertainties. Typical values are $0.05$~dex on [M/H], $0.05$~mag on E(B$-$V), and $0.10$~mag on DM. A flat prior is adopted for the age over the full range of the isochrone grid (6--15~Gyr). For each GC, two independent fitting runs were performed on the ($m_{F606W}$, $m_{F606W}-m_{F814W}$) and ($m_{F814W}$, $m_{F606W}-m_{F814W}$) CMDs. The final parameter estimates are taken as the average of the two runs, with uncertainties given by the combined 16th and 84th percentiles of the posterior distributions.

\section{Analytical modelling of the AMRs}\label{sec:amr_modelling}

To quantify the evolution of each progenitor and to resolve the membership of dynamically ambiguous clusters, we modelled the observed AMRs using the simple leaky-box chemical-evolution framework discussed by \citet{leaman13}, as first introduced in the CARMA series in \citetalias{aguado25} and following the analytical description of \citet{prantzos08}. For a system evolving with a constant SFR, in which metal loss is absorbed into an effective yield and the gas reservoir declines until star formation ends at $t_f$, the metallicity is related to the look-back time of formation as
\begin{equation}\label{eq:amr}
    t([\mathrm{M/H}]) = t_f + \Delta t \cdot \exp\!\left(\frac{-10^{[\mathrm{M/H}]}}{p}\right),
\end{equation}
where $p$ is the effective chemical yield relative to $Z_\odot$, $\Delta t = t_i - t_f$ is the duration of star formation, $t_i$ is the look-back time at the onset of star formation, and $t_f$ is the look-back time at which star formation ends (which we take as the galaxy's accretion time onto the MW). The yield parameter $p$ is proportional to the total initial stellar mass, $M_\star$, of the progenitor galaxy through the empirical calibration
\begin{equation}\label{eq:p_mass}
    p = 0.005 \left(\frac{M_\star}{10^6\,\mathrm{M}_\odot}\right)^{0.4},
\end{equation}
following \citet{prantzos08} and the empirical dwarf-galaxy scaling determined by \citet{dekelwoo03}. Inverting Eq.~\ref{eq:p_mass}, the initial stellar mass is $M_\star = 10^6\,(p/0.005)^{2.5}\,\mathrm{M}_\odot$.

For each progenitor, we fitted Eq.~\ref{eq:amr} to the observed $([\mathrm{M/H}],\,t)$ values using an MCMC sampler implemented with \texttt{emcee} \citep{foreman-mackey2013emcee}. The model has five free parameters: $(p,\,t_i,\,t_f,\,\eta_{\rm low},\,\eta_{\rm high})$, where $\eta_{\rm low}$ and $\eta_{\rm high}$ are the logarithms of the asymmetric intrinsic scatter in age (below and above the model curve, respectively). The log-likelihood was computed as a sum of asymmetric Gaussian terms in both age and metallicity, accounting for the asymmetric observational errors from the isochrone fitting. A log-normal prior was adopted for $M_\star$, centred on the literature estimates for the stellar mass of each system: $\sim5\times10^8$\,M$_\odot$ for GSE \citep{helmi18}, $\sim10^8$\,M$_\odot$ for H99 \citep{koppelman19}, $\sim10^8$\,M$_\odot$ for Sequoia \citep{myeong19}, and $\sim10^9$\,M$_\odot$ for Sgr \citep{vasiliev21sgr}. We set a weakly informative width of $\sigma_{\ln M} = 1.00$ to allow the likelihood to drive the fit while preventing exploration of unphysical masses. The temporal parameters have a joint flat prior over the physically ordered domain $8 < t_f < t_i < 15$\,Gyr for GSE and Seq, $5 < t_f < t_i < 15$\,Gyr for Sgr, and $8 < t_f < t_i < 20$\,Gyr for H99. Finally, the intrinsic age scatters, $\sigma_{\rm int,low}=\exp(\eta_{\rm low})$ and $\sigma_{\rm int,high}=\exp(\eta_{\rm high})$, were assigned half-normal shrinkage priors with a scale of 0.18\,Gyr (0.10\,Gyr for H99), implemented in log-scatter space with the corresponding Jacobian. Throughout this work, all quoted parameter values correspond to the median of the posterior distribution, which may not coincide with the peak of the marginalised distribution (mode) when posteriors are asymmetric, but is preferred for its robustness to skewness, and for consistency with previous CARMA works.

As discussed by \citet{dekelwoo03} and \citet{prantzos08}, this leaky-box model provides a fair representation of the early chemical evolution of dwarf galaxies that evolve approximately in isolation before their star and GC formation is truncated by a physical process, in our case accretion onto the MW. The parameter $p$ must therefore be interpreted as an effective yield: it incorporates metal loss through outflows and is connected to stellar mass through the empirical calibration of Eq.~\ref{eq:p_mass}, rather than representing a universal nucleosynthetic yield. Likewise, $t_i$ and $t_f$ are effective descriptors of the cluster-forming epoch and of each progenitor's accretion time, and are used as differential diagnostics within the homogeneous CARMA age scale. With this interpretation, our relative conclusions about progenitor masses and the accretion sequence are robust. It should be noted that for the systems constrained by only a few clusters (H99, Seq, and Sgr), the posterior masses remain strongly informed by their literature-based priors and their agreement with published values is therefore a consistency check rather than an independent confirmation.

\section{Results}\label{sec:results}

\subsection{Isochrone-fitting results}

The results of the isochrone fitting for all 24 GCs in our sample are presented in Table~\ref{tab:results}, which lists the best-fit values of global metallicity [M/H], colour excess E(B$-$V), DM, and age for each cluster, together with the adopted progenitor association. When considering the whole sample, the resulting relative ages span a wide range, from $8.07$~Gyr (Terzan~7, Sgr) to $14.16$~Gyr (Terzan~8, Sgr), with the full sample covering approximately 6~Gyr of cosmic history. The GSE clusters in our sample are among the oldest, with ages predominantly in the range $12$--$14$~Gyr, consistent with the early assembly of the GSE dwarf galaxy. The Sgr system shows the largest internal age dispersion of all four progenitors, with the younger clusters Pal~12 and Terzan~7 at significantly higher metallicities compared to the old Terzan~8 and Arp~2, reflecting the extended and metal-enriched SFH expected for this ongoing merger event. The H99 and Seq clusters are predominantly old ($>11$~Gyr). All age values reported in this work are to be interpreted strictly in a relative sense within the homogeneous CARMA framework, as discussed in \citetalias{massari23}.

\subsection{Consistency checks}

\begin{figure*}[ht!]
\sidecaption
\begin{minipage}[b]{12cm}
        \includegraphics[width=\linewidth]{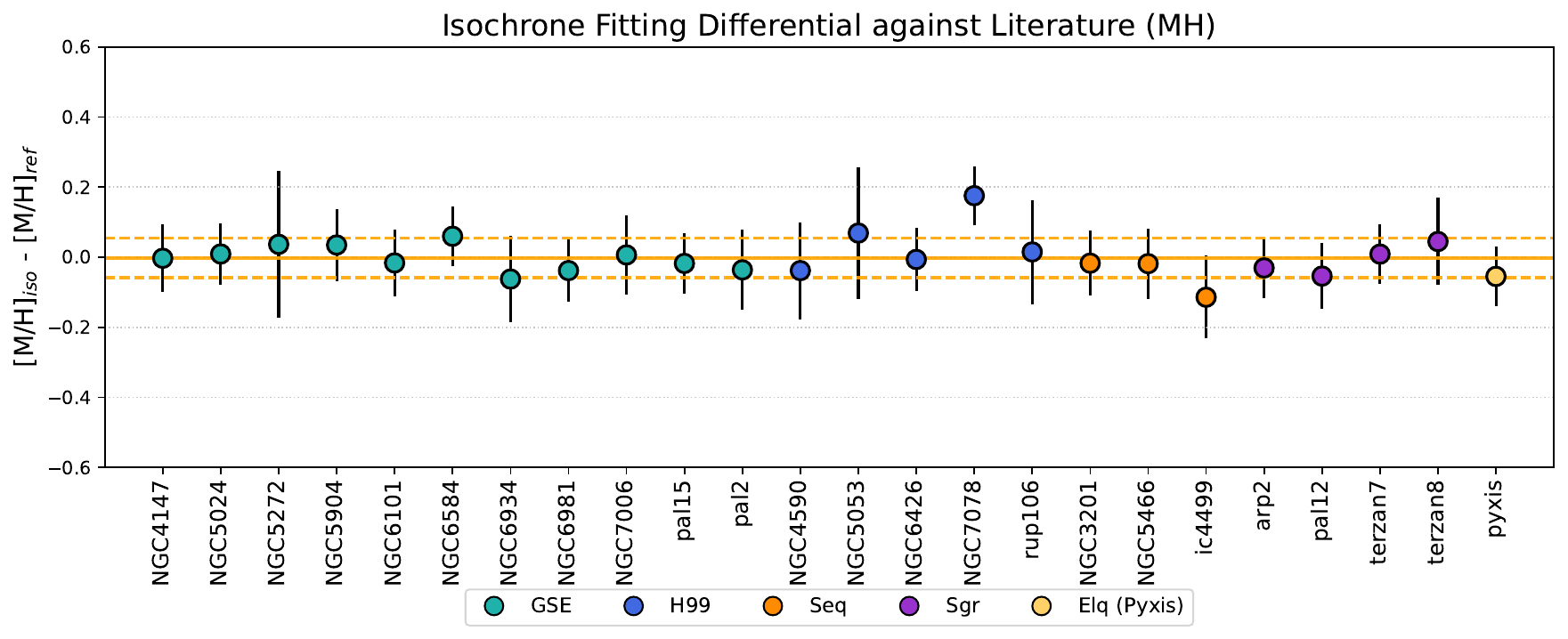}\\
        \includegraphics[width=\linewidth]{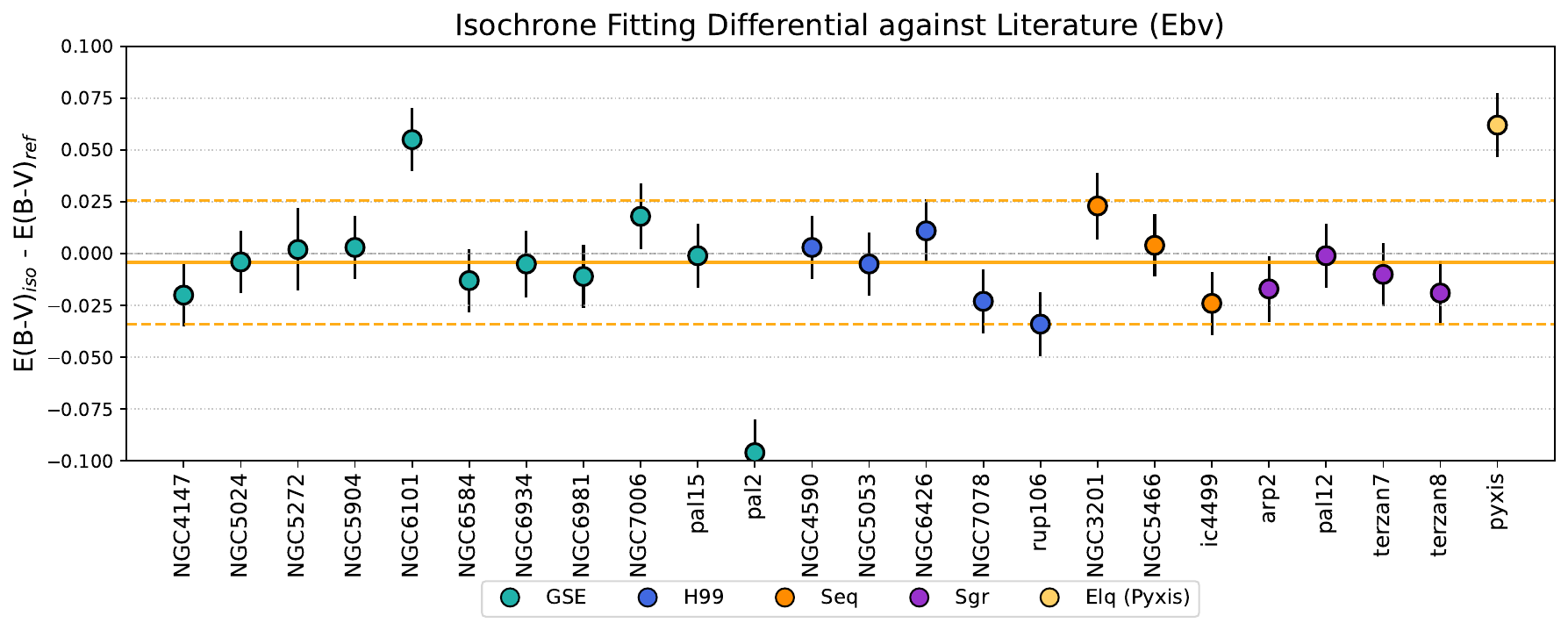}\\
        \includegraphics[width=\linewidth]{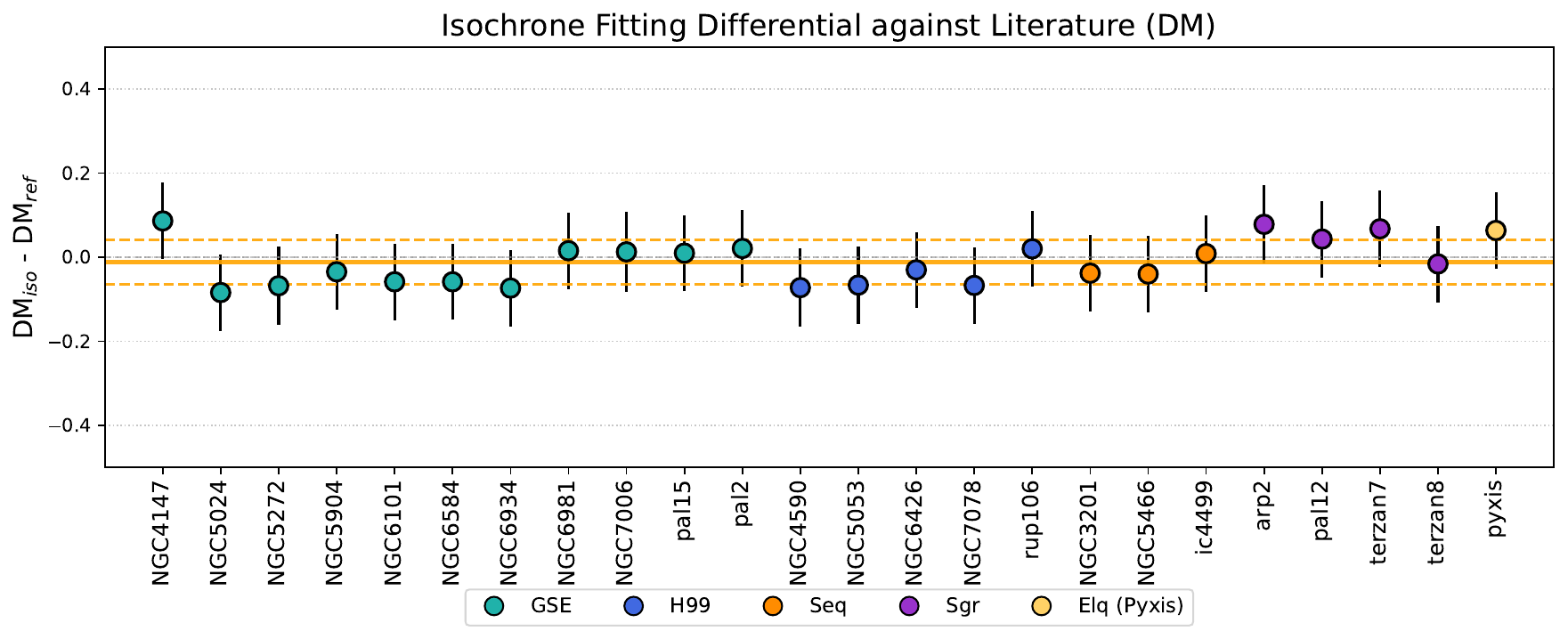}
\end{minipage}
\caption{Difference between the isochrone-fitting output values and the reference literature values for global metallicity (\textit{top}), colour excess (\textit{middle};  \citealt{harris10}), and true DM (\textit{bottom}; \citealt{baumgardt21}). All 24 GCs are included and sorted horizontally by their associated progenitor system. Coloured symbols follow the progenitor colour scheme adopted throughout this work. The dashed horizontal grey line marks zero difference. }
\label{fig:diff}
\end{figure*}

As a consistency test of the parameters derived from the isochrone-fitting procedure, we compared the output values of [M/H] and E(B$-$V) with the reference value listed in the \citet{harris10} catalogue, and the true DM with the independent, highly accurate determinations from \citet{baumgardt21}. The results of this comparison are shown in Fig.~\ref{fig:diff}, where all 24 GCs are displayed grouped by progenitor system. For Terzan~7, we adopted the spectroscopic value $[\mathrm{Fe/H}]=-0.59$ from \citet{sbordone05} rather than the calcium-triplet-based measurements originally provided by \citet{dacosta95}.

The fitted global metallicities (top panel) show a good agreement with the catalogue iron abundances converted to [M/H] following the prescription of \citet{salaris93} and assuming the [$\alpha$/Fe]-[Fe/H] relation for accreted systems adopted in \citetalias{massari23}. The mean difference is consistent with zero within the uncertainties, and the scatter around zero is remarkably small ($<0.05$). The colour excess values (middle panel) are well recovered for most of the sample; a few outliers are caused by clusters with high extinction (e.g. Pal~2 with $E(B-V) \approx 1.14$~mag). In fact, as noted in \citet{massari26}, most of the Harris $E(B-V)$ estimates come from isochrone fits that provide distance and reddening altogether, whereas we prefer to use \citet{baumgardt21} as a reference for distance. This means that there exist clusters with DM in good agreement with the \citet{baumgardt21} compilation, but more discrepant compared to Harris, and hence with its reddening values. The distance moduli (bottom panel) are in excellent agreement with the \citet{baumgardt21} determinations, confirming the reliability of our adopted distance scale. Overall, the comparison confirms the reliability of the CARMA isochrone-fitting framework when applied to this heterogeneous sample of accreted GCs, and demonstrates that no systematic bias is introduced by the use of two different input photometric catalogues (HUGS and Sarajedini et al. 2007).

\subsection{Membership assignments}\label{sec:membership}

Of the 24 GCs with new age determinations in this work (Table~\ref{tab:results}), 9 have unambiguous dynamical associations (4 to Sgr, 4 to GSE, and 1 to Elqui), while the remaining 15 have contested origins between GSE and a second candidate progenitor, with their integrals of motion placing them at the boundary between two systems \citep{callingham22, chen24, massari26}. To resolve the ambiguous cases, we leveraged our age estimates in a two-step procedure. First, we constructed reference AMR curves for each candidate progenitor. For GSE, we fitted Eq.~\ref{eq:amr} using a kinematically unambiguous reference sample composed of the 4 GSE clusters whose dynamical membership is uncontested in our sample (NGC~4147, NGC~6981, Pal~2, and Pal~15), and of the 11 clusters chrono-dynamically associated with GSE by \citet{massari26}, giving 15 clusters in total. This preliminary MCMC run yields a robust best-fit GSE model, free from contamination by uncertain members. For H99, Seq, and the high-energy group (H-E), we adopted reference AMR curves with parameters from the literature \citep{kruijssen19, myeong19, helmi20}, as the number of unambiguously associated clusters in our sample is too small to constrain independent fits.

After each progenitor AMR model was determined, for each of the 15 clusters with uncertain dynamical assignments, we generated 5000 Monte Carlo realisations of their position in the $([\mathrm{M/H}],\,t)$ plane, sampling from asymmetric Gaussians defined by the isochrone-fitting results. For each realisation, we determined the Mahalanobis distance from the model of the candidate progenitor, $j$. To do so, we computed the metallicity and age residuals of each cluster realisation from each AMR curve sampling point, $s$\footnote{Each curve is sampled by 2000 points}, as  $\Delta Z_j(s)=[\mathrm{M/H}]-[\mathrm{M/H}]_j(s)$ and $\Delta t_j(s)=t-t_j(s)$, respectively. We then defined the squared Mahalanobis distance to progenitor $j$ as the minimum over all the sampled curve points,
$d_{{\rm M},j}^2=\min_s\left[(\Delta Z_j(s)/\sigma_{Z,{\rm err}})^2+\Delta t_j(s)^2/(\sigma_{t,{\rm err}}^2+\sigma_{{\rm int},j}^2)\right]$. Here $\sigma_{Z,{\rm err}}$ and $\sigma_{t,{\rm err}}$ are the observational uncertainties on the individual cluster metallicity and age, and $\sigma_{{\rm int},j}$ is the intrinsic age spread associated with the reference AMR.
Because the candidate AMRs might have different intrinsic spreads, following the covariance-aware Gaussian treatment of two-dimensional relations with intrinsic scatter described by \citet{hogg10}, we used the normalised statistic $Q_j=d_{{\rm M},j}^2+\ln|\boldsymbol{\Sigma}_j|$, where
$\boldsymbol{\Sigma}_j={\rm diag}(\sigma_{Z,{\rm err}}^2,\,\sigma_{t,{\rm err}}^2+\sigma_{{\rm int},j}^2)$. This way, the determinant term prevents a broader AMR from being artificially favoured in the statistical assignments. For GSE, the posterior median intrinsic spread inferred from the initial MCMC AMR fit is $\sigma_{{\rm int},GSE}=0.22$\,Gyr. For the other disputed progenitors, we used the intrinsic spreads inferred from their corresponding MCMC AMR fits described in Sect.~\ref{sec:amr_modelling}. Since GSE is always one of the candidate progenitors -- i.e. each of the 15 contested clusters is disputed between GSE and one alternative system -- we computed the pairwise probability as
\begin{equation}\label{eq:pgse}
    P(\mathrm{GSE}) = \frac{N\{Q_{\rm GSE} < Q_{\rm alt}\}}{N_{\rm MC}}
,\end{equation}
where $N_{\rm MC} = 5000$. A value $P(\mathrm{GSE}) \gg 0.5$ indicates unambiguous GSE membership in the AMR plane, while $P(\mathrm{GSE}) \ll 0.5$ favours the alternative progenitor. However, this single probability does not account for the uncertainty in the AMR models. We therefore repeated the above calculation 1000 times, each time drawing the GSE curve parameters and the intrinsic widths of both candidate AMRs from their posterior distributions, while retaining the literature-based curve parameters for the alternative progenitor. For each draw we recomputed $P(\mathrm{GSE})$ from the same 5000 Monte Carlo realisations. To quantify the significance of each association, we then defined $N_\sigma=|P_{50}-0.5|/\sigma_P$, where $\sigma_P=(P_{84}-P_{16})/2$, and $P_{16}$, $P_{50}$, and $P_{84}$ are the corresponding percentiles across the posterior distribution of the 1000 $P(\mathrm{GSE})$ values. We adopted a criterion, $N_\sigma>2$, to associate a GC with one progenitor unambiguously.

This first step resolved the debated associations of several GCs. Hence, we repeated the same procedure by updating the GSE AMR model with the newly confirmed members, while retaining the literature-based curves for the other progenitors. The resulting P(GSE) and $N_\sigma$ are quoted in Table \ref{tab:probabilities} and illustrated in Fig.~\ref{fig:membership_prob} for the case of the GSE versus H99 discrimination.
After this last procedure, of the 15 clusters with ambiguous dynamical assignments, the method robustly assigns 7 to GSE and 3 to H99 (NGC~4590, NGC~5053, and Rup~106). The remaining 5 clusters remain uncertain: NGC~3201 ($N_\sigma=0.07$), NGC~5466 ($N_\sigma=0.77$), IC~4499 ($N_\sigma=1.95$), NGC~6426 ($N_\sigma=1.45$), and NGC~7078 ($N_\sigma=1.10$). For these ambiguous cases, we maintain the preferred dynamical association from \citet{massari25}: NGC~3201, NGC~5466, and IC~4499 are assigned to Seq, while NGC~6426 and NGC~7078 are assigned to H99. NGC~7078 nominally favours GSE, but since the preference is not statistically significant and its prograde and relatively high circularity orbit favours H99, we retain the H99 association. Throughout this work, their memberships are labelled with the double notation (e.g. Seq--GSE, H99--GSE) to reflect the residual uncertainty. The final assignments are also reported in Table~\ref{tab:results}.

\begin{table*}[t]
\centering
\begin{minipage}[c]{0.45\textwidth}
\centering
\caption{\label{tab:probabilities} Median pairwise membership probabilities for the 15 kinematically ambiguous clusters.}
\small
\setlength{\tabcolsep}{3pt}
\begin{tabular}{@{}lcccc@{}}
\hline \hline
Cluster & Alternative & $P_{50}$(GSE) & $N_\sigma$ & Assigned \\
\hline
NGC~3201 & GSE--Seq & 0.505 & 0.07 & Seq--GSE \\
NGC~4590 & GSE--H99 & 0.038 & $>$10 & H99 \\
NGC~5024 & GSE--H99 & 1.000 & $>$10 & GSE \\
NGC~5053 & GSE--H99 & 0.308 & 2.99 & H99 \\
NGC~5272 & GSE--H99 & 0.870 & $>$10 & GSE \\
NGC~5466 & GSE--Seq & 0.403 & 0.77 & Seq--GSE \\
NGC~5904 & GSE--H99 & 1.000 & $>$10 & GSE \\
NGC~6101 & GSE--Seq & 0.999 & $>$10 & GSE \\
NGC~6426 & GSE--H99 & 0.358 & 1.45 & H99--GSE \\
NGC~6584 & GSE--H99 & 1.000 & $>$10 & GSE \\
NGC~6934 & GSE--H-E & 1.000 & $>$10 & GSE \\
NGC~7006 & GSE--Seq & 1.000 & $>$10 & GSE \\
NGC~7078 & GSE--H99 & 0.611 & 1.10 & H99--GSE \\
IC~4499 & GSE--Seq & 0.317 & 1.95 & Seq--GSE \\
Rup~106 & GSE--H99 & 0.272 & 5.90 & H99 \\
\hline
\end{tabular}
\tablefoot{$P_{50}$(GSE) was computed from the normalised Mahalanobis likelihood (Eq.~\ref{eq:pgse}) against the alternative progenitor listed in the second column. Ambiguity was assessed from $N_\sigma$, the posterior-propagated significance of the offset of $P_{50}(\mathrm{GSE})$ from $0.5$ (Sect.~\ref{sec:membership}), rather than from a fixed probability window. A cluster association was considered unsolved if $N_\sigma < 2$.}
\end{minipage}\hfill
\begin{minipage}[c]{0.50\textwidth}
\centering
\includegraphics[width=\linewidth]{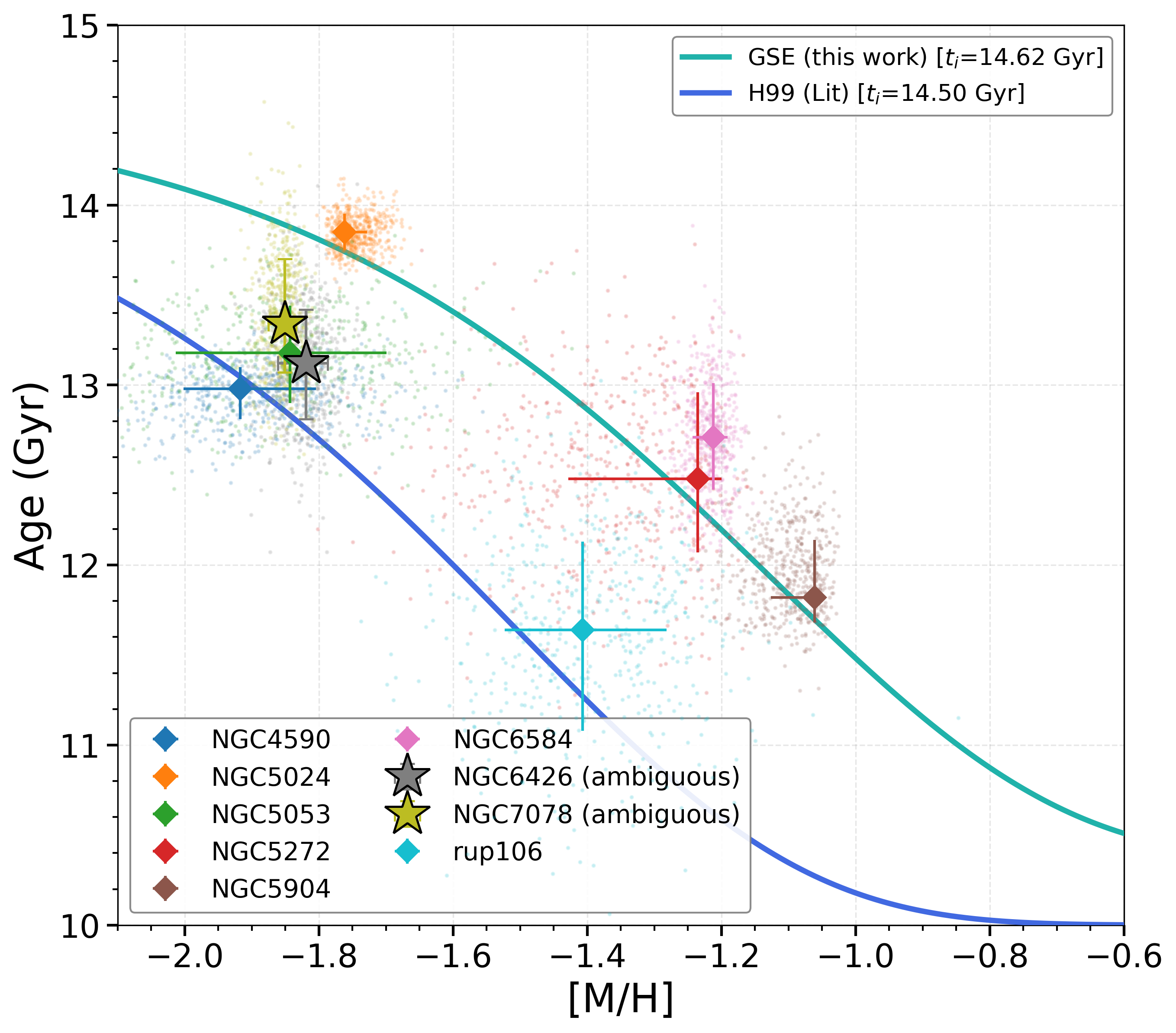}
\captionof{figure}{Pairwise membership discrimination between the GSE and H99 AMR models (step~2). Each cluster tested for GSE--H99 ambiguity is shown with its best-fit age and metallicity (diamond) and the corresponding Monte Carlo realisations (coloured dots), drawn from the observational uncertainties. The solid curves show the best-fit GSE (teal) and H99 (blue) AMR models. NGC~6426 and NGC~7078, marked with stars, are the two clusters in this panel whose classification remains statistically ambiguous ($N_\sigma < 2$).}\label{fig:membership_prob}
\end{minipage}
\end{table*}

\begin{table}[h!]
\centering
\caption{\label{tab:final_amr_params} Parameter estimates for the final MCMC AMR models.}
\small
\setlength{\tabcolsep}{4pt}
\begin{tabular}{lcccc}
\toprule
System & $p$ & $t_i$ & $t_f$ & $\text{M}_\star$ \\
 & & [Gyr] & [Gyr] & [$10^8 \text{M}_\odot$] \\
\midrule
GSE & $0.075_{-0.019}^{+0.027}$ & $14.62_{-0.22}^{+0.21}$ & $10.36_{-0.87}^{+0.64}$ & $8.61_{-4.36}^{+9.74}$ \\[1ex]
H99 & $0.026_{-0.009}^{+0.013}$ & $14.36_{-0.71}^{+1.12}$ & $11.29_{-1.21}^{+0.90}$ & $0.61_{-0.39}^{+1.01}$ \\[1ex]
Sgr & $0.080_{-0.010}^{+0.013}$ & $14.83_{-0.27}^{+0.13}$ & $8.14_{-0.11}^{+0.24}$ & $10.35_{-3.02}^{+4.59}$ \\[1ex]
Seq & $0.028_{-0.008}^{+0.012}$ & $14.45_{-0.57}^{+0.38}$ & $11.70_{-0.71}^{+0.49}$ & $0.77_{-0.45}^{+1.11}$ \\
\midrule
LKH$^{\ast}$ & $0.066_{-0.019}^{+0.028}$ & $14.52_{-0.26}^{+0.27}$ & $12.19_{-0.59}^{+0.49}$ & $6.30_{-3.60}^{+8.95}$ \\
\bottomrule
\end{tabular}
\tablefoot{Quoted uncertainties are asymmetric $1\sigma$ errors. $^{(\ast)}$The LKH parameters are taken from \citet{massari26}.}
\end{table}

\subsection{AMR-based characterisation of the merger events}\label{sec:mcmc_results}

\begin{figure*}[ht!]

\centering
\includegraphics[width=\textwidth]{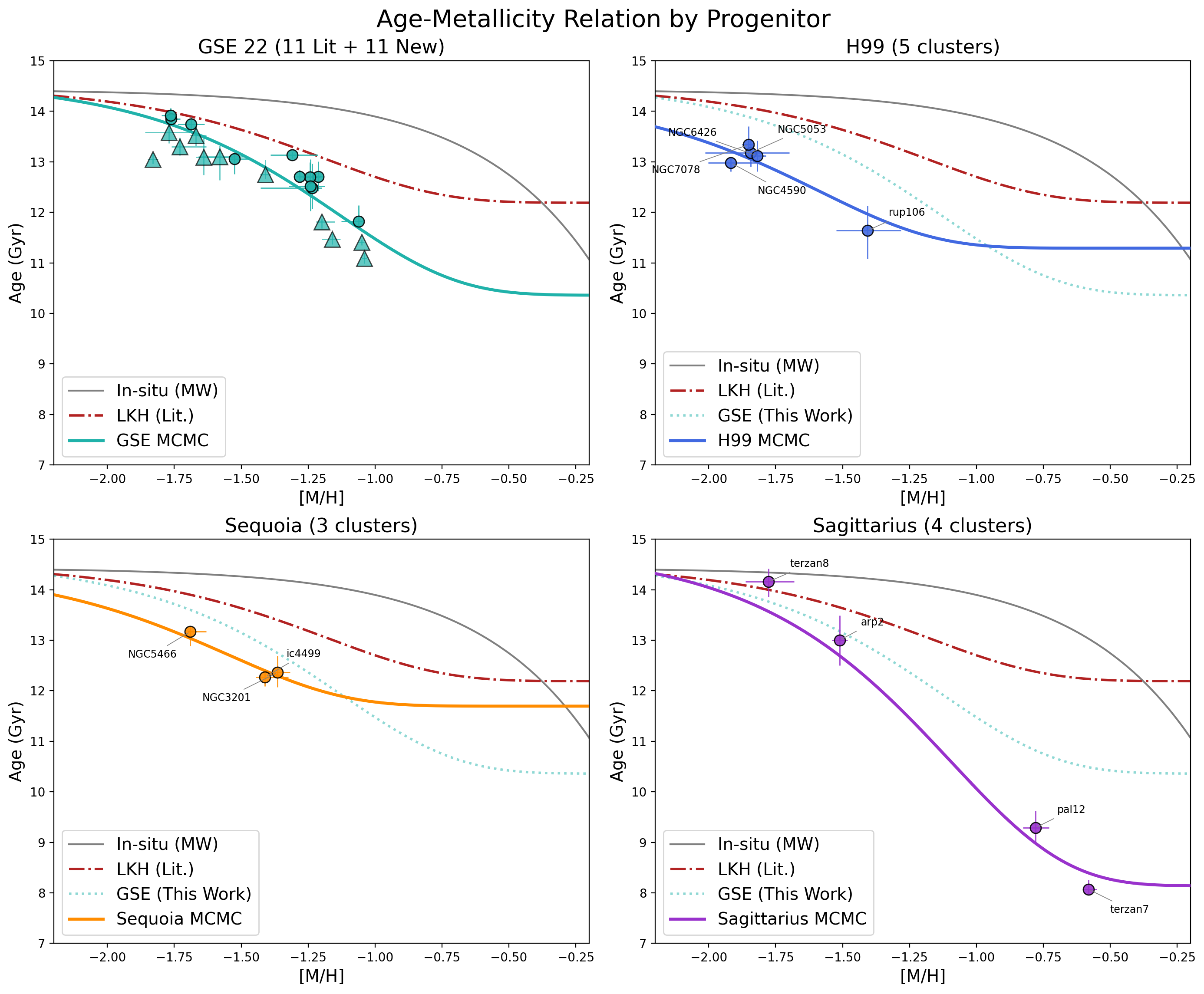}
\caption{AMRs for the four progenitor systems: GSE (22 clusters), H99 (5), Sagittarius (4), and Sequoia (3). Filled circles mark the 24 GCs with new age determinations from this work; triangles denote the 11 GSE literature clusters from \citetalias{aguado25} and \citet{massari26}. The solid curves show the best-fit MCMC AMR models from Table~\ref{tab:final_amr_params}. The grey curve in each panel shows the in situ MW sequence from \citetalias{zerbinati26}, and the dash-dotted dark red curve shows the LKH AMR from \citet{massari26}, both plotted as reference.}
\label{fig:grid_amr_4panel}
\end{figure*}

Once the ambiguous cases of membership have been solved, the observed AMRs for the four progenitor systems are presented in Fig.~\ref{fig:grid_amr_4panel}. The four systems trace distinct loci in the age-metallicity plane, reflecting their individual SFHs, as discussed below.

We fit each observed AMR with the MCMC algorithm described in Sect.~\ref{sec:amr_modelling}. The best-fit models are shown in Figs.~\ref{fig:mcmc_amr_models_1} and \ref{fig:mcmc_amr_models_2}, and the resulting parameters are summarised in Table~\ref{tab:final_amr_params}. For each system, the left panel shows the posterior distributions of the five model parameters, and the right panel shows the best-fit AMR curve overlaid on the cluster data. We discuss the results for each progenitor in turn. We stress that all ages in this work are to be interpreted in a relative sense \citep[see][]{massari23}: the absolute values of $t_f$ reported below should therefore not be taken at face value, but rather used as differential diagnostics of the accretion sequence.

\textit{Gaia-Sausage-Enceladus.} The GSE system was fitted using all 11 confirmed GSE members in our sample, supplemented by the 11 literature clusters from \citetalias{aguado25} and \citet{massari26}. The resulting yield parameter is $p = 0.075^{+0.027}_{-0.019}$, corresponding to a total initial stellar mass of $M_\star = (8.61^{+9.74}_{-4.36})\times 10^8\,\mathrm{M}_\odot$. This is in good agreement within the uncertainties with the independent estimate of $5\times10^8\,\mathrm{M}_\odot$ from \citet{helmi18}, as well as with slightly more massive values such as those of \citet{vincenzo19}, \citet{das20}, and \citet{sante26}. Globular cluster formation in the GSE progenitor was terminated by accretion onto the MW at $t_f = 10.36^{+0.64}_{-0.87}$~Gyr, consistent with the timing of $\sim10$~Gyr derived from kinematics \citep{helmi18, belokurov18} and from the analysis of field SFHs \citep{gallart19, gonzalezkoda25}.
The total duration of GC formation in GSE ($\sim4.3$~Gyr) reflects a sustained but eventually quenched SFH \citep{valenzuela24, lardo26}. 

\textit{Helmi streams (H99).} First, we note that although NGC~7078 was previously classified as a GSE cluster in \citetalias{aguado25}, its AMR-based preference is not statistically significant, while its prograde orbit and relatively high circularity ($circ\sim0.6$; \citealt{massari25}) favours an H99 association. The total sample of five H99 clusters (NGC~4590, NGC~5053, NGC~6426, Rup~106, and NGC~7078) defines a compact AMR with a significantly lower yield than GSE, $p = 0.026^{+0.013}_{-0.009}$, and a correspondingly lower stellar mass of $M_\star = (0.61^{+1.01}_{-0.39})\times 10^8\,\mathrm{M}_\odot$. This places the H99 progenitor firmly in the regime of a low-mass dwarf galaxy, consistent with kinematic estimates of $\sim10^8\,\mathrm{M}_\odot$ from \citet{koppelman19}. The H99 system was accreted at $t_f = 11.29^{+0.90}_{-1.21}$~Gyr, nominally about 1.0~Gyr before the GSE merger, although the two accretion times are consistent within their $1\sigma$ uncertainties. This is in good agreement with the simulation-based inference by \citet{sante26}, who predict a lookback infall time of $10.1^{+0.7}_{-0.9}$~Gyr for the H99 progenitor. We note that a direct comparison between ours and the accretion time by \citet{sante26} requires one to account for the 0.65~Gyr zero-point offset between the CARMA relative age scale and the absolute lookback times calibrated via asteroseismology in \citetalias{zerbinati26}; once this offset is applied, our $t_f$ translates to an absolute value of $\sim$10.6~Gyr, fully consistent with \citet{sante26}. The large uncertainties on $t_f$ reflect the limited size of the H99 GC sample and the spread of the cluster positions around the model curve.

\textit{Sagittarius.} The four Sgr clusters span the widest range in both age and metallicity of the four progenitors, and the MCMC returns the broadest AMR of the sample. The yield parameter $p = 0.080^{+0.013}_{-0.010}$ is the highest of the four systems, yielding the largest stellar mass, $M_\star = (10.35^{+4.59}_{-3.02})\times 10^8\,\mathrm{M}_\odot$, in good agreement with the estimate of $\sim10^8$--$10^9\,\mathrm{M}_\odot$ from N-body models of the Sgr disruption \citep{vasiliev21sgr}, and with the simulation-based inference by \citet{sante26}, who predict $\log(M_\star/\mathrm{M}_\odot) = 8.8^{+0.2}_{-0.2}$.
The extended SFH of Sgr reflects its late accretion time and its high stellar mass, which allowed it to sustain star formation well after its accretion started \citep{hughes19, zerbinati26}. According to our AMR fit, Sgr was accreted $t_f=8.14^{+0.24}_{-0.11}$ Gyr ago. This value is in good agreement with other estimates obtained from the Sgr GC system \citep[e.g.][]{kruijssen20, lardo26}, and with the simulation-based determination by \citet{sante26}. However, we remark that for a merger event this late, our assumption that the accretion time coincides with the time at which GC formation ends is likely inaccurate. Sgr is notoriously still merging with the MW, and this explains why our accretion time is earlier compared to other literature estimates coming from stellar dynamics~\citep{vasiliev21sgr} or SFH-based \citep{ruizlara20} arguments. More likely, for Sgr it is more reasonable to think of our $t_f$ as related to one of its first pericentric passages around the MW.

\textit{Sequoia.} With only three adopted Seq members (NGC~3201, NGC~5466, and IC~4499), the MCMC provides the most uncertain parameter estimates of the four systems. The low yield, $p = 0.028^{+0.012}_{-0.008}$, and the inferred stellar mass, $M_\star = (0.77^{+1.11}_{-0.45})\times 10^8\,\mathrm{M}_\odot$, place Seq among the lowest-mass progenitors. Globular cluster formation was brief ($\Delta t \approx 2.8$~Gyr) and restricted to the metal-poor regime, consistent with the slow chemical evolution and early quenching expected for a galaxy of this stellar mass \citep{myeong19, feuillet21, naidu22, dodd25}. The accretion time of Seq ($t_f \approx 11.70$~Gyr) is nominally earlier than H99 by $\sim$0.4~Gyr and than GSE by $\sim$1.4~Gyr, placing Sequoia as the first accretion event to follow LKH in the MW assembly history. Both the stellar mass and the accretion time are consistent with the simulation-based inference by \citet{sante26}, who predict $\log(M_\star/\mathrm{M}_\odot) = 7.4^{+0.5}_{-0.4}$ and a lookback infall time of $11.3^{+0.6}_{-0.5}$~Gyr for the Sequoia progenitor (our $t_f$ translates to $\sim$11.1~Gyr once the 0.65~Gyr absolute zero-point offset from \citetalias{zerbinati26} is applied).

It is worth noting that the AMRs of the Helmi streams and Sequoia are remarkably similar in both slope and metallicity range, with comparable yields ($p_{\rm H99} = 0.026$, $p_{\rm Seq} = 0.028$) and stellar masses. This similarity implies that, in the absence of dynamical information, their cluster members would be virtually indistinguishable in the age-metallicity plane alone. The reliable separation of these two systems therefore relies critically on their distinct orbital properties: H99 clusters follow prograde, moderately inclined orbits, while Seq clusters occupy retrograde, high-energy orbits \citep{koppelman19, myeong19}.

\subsection{Field star comparison}

\begin{figure}[ht!]
\center{
\includegraphics[width=\columnwidth]{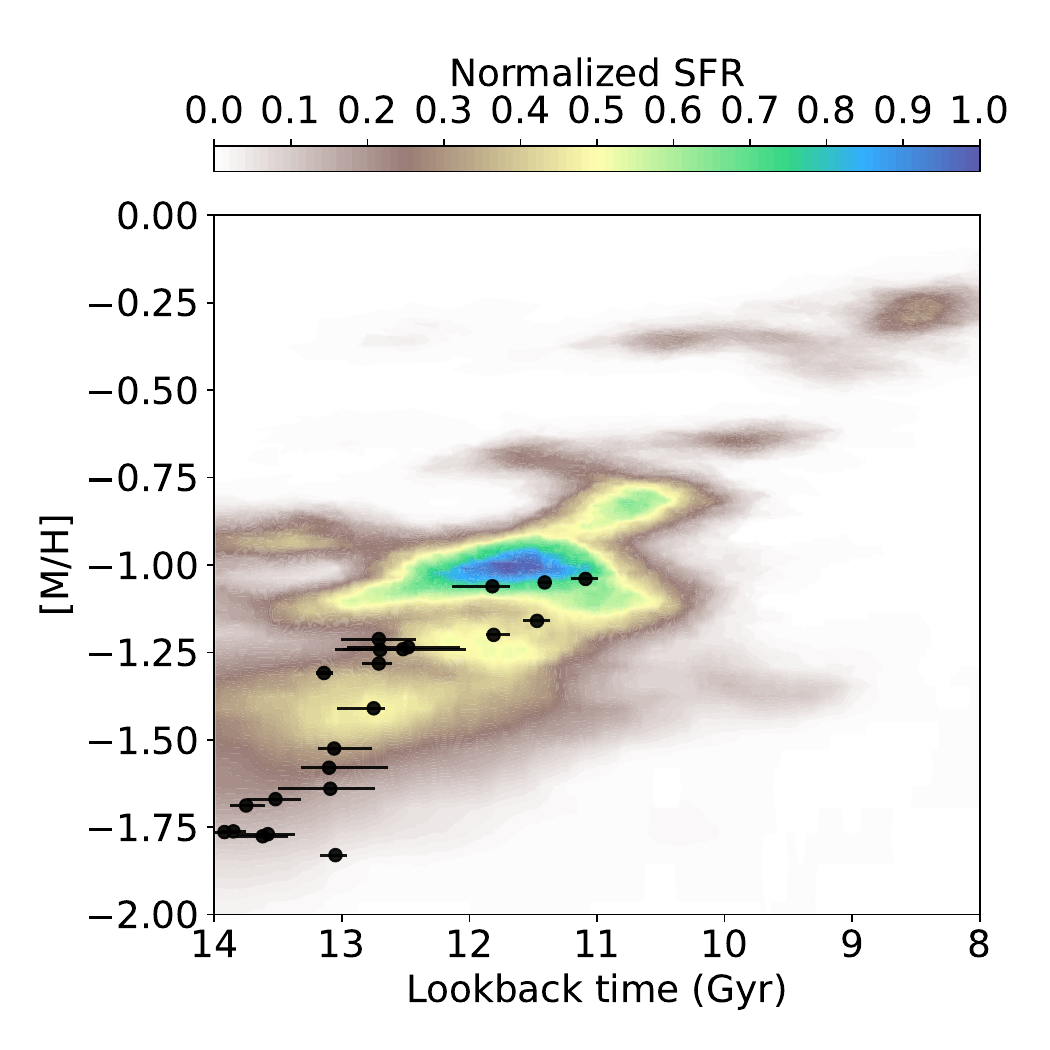}
}
\caption{Comparison between the AMR of the GSE GCs from this work and the dynamically evolved SFHs of GSE field stars.}
\label{fig:age_fieldstars}
\end{figure}

An independent consistency check on the progenitor assignments and the AMR models can be obtained by comparing the positions of the GCs in the age-metallicity plane with the distribution of halo field stars sharing the same dynamical origin. Figure~\ref{fig:age_fieldstars} superimposes the AMR positions of the GSE GCs studied in this work on the equivalent derived from the SFH of GSE field stars obtained via the CMD-fitting technique of \citep{gonzalezkoda25}. When considering the possible systematics at play (see the appendix of the same paper), the GC sequence closely follows the ridge line of the stellar density distribution, providing independent support for the progenitor assignments derived from the MCMC modelling and confirming that the GCs are faithful tracers of the chemical enrichment history of their parent galaxy.

\subsection{Comparison between GSE and the LMC}

\begin{figure}[ht!]
\center{
\includegraphics[width=\columnwidth]{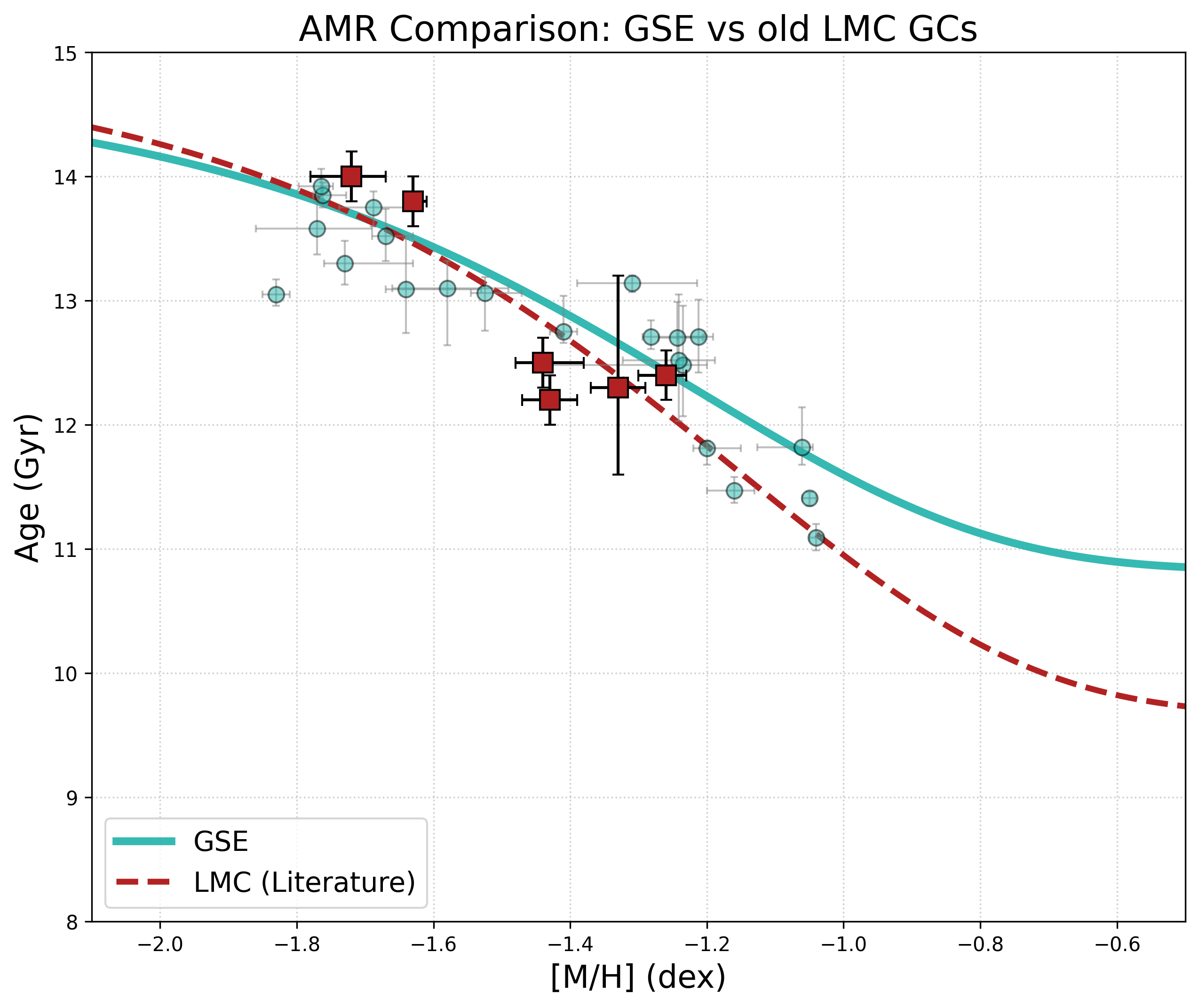}
}
\caption{Comparison between the AMRs of GSE GCs and old LMC GCs studied in \citetalias{niederhofer25}. The similarity between the two sequences suggests analogous early star-formation efficiencies in the two progenitor systems.}
\label{fig:gse_lmc_comparison}
\end{figure}

The CARMA series recently extended its homogeneous isochrone-fitting framework to seven old star clusters in the LMC (\citetalias{niederhofer25}), providing a unique direct comparison for the AMR of an isolated dwarf galaxy with an independently constrained SFH. In order to ensure a physically meaningful comparison with GSE, we restricted the LMC sample to six old clusters ($\gtrsim 11$~Gyr), which trace the early, pre-interaction chemical evolution of the LMC, and we excluded the likely accreted cluster NGC~1841. The younger LMC cluster population ($\lesssim 10$~Gyr) was deliberately excluded for two reasons: (i) these clusters post-date the GSE merger event, and therefore probe a fundamentally different evolutionary epoch, preventing any direct comparison; and (ii) their formation is known to have been significantly enhanced by tidal interactions between the LMC and its satellite, the Small Magellanic Cloud \citep[SMC;][]{harris09, piatti19}, which introduces an environmental effect that would break the analogy with the self-enrichment history of GSE prior to its accretion onto the MW. The six old LMC clusters thus provide the most homogeneous and physically motivated comparison sample for evaluating the early star-formation efficiency of the two systems on an equal footing.

In Fig.~\ref{fig:gse_lmc_comparison} we compare the AMR of the confirmed GSE clusters from this work with the six old LMC clusters from \citetalias{niederhofer25}. The two sequences trace strikingly similar loci in the age-metallicity plane, both displaying old ($\gtrsim 11$~Gyr), predominantly metal-poor ($-2.0 \lesssim \mathrm{[M/H]} \lesssim -1.0$~dex) clusters with comparable slopes. This convergence suggests that the GSE and LMC progenitors experienced analogous early star-formation efficiencies and chemical enrichment timescales, despite their very different accretion histories: the GSE was fully disrupted $\sim10$~Gyr ago, while the LMC is still on its first infall into the MW \citep{besla10}.  Interestingly, with a stellar mass of $\simeq 2.5\times 10^9$ \citep{Besla2012_MCs} acquired by the LMC over the course of its whole 13.5 Gyr evolution, assuming an approximately constant average SFR the mass that the LMC would have formed during its first 3.5 Gyr of evolution is very much similar to the one inferred for GSE in this study. We are therefore in front of two systems relatively similar in origin but with a completely different fate. One difference, however, may be in its GC population: while the LMC has 15 known old GCs (older than $\sim$8 Gyr), the GC population of GSE is larger, probably reflecting a more intense initial star formation epoch consistent with it being formed in a denser environment \citep{Gallart2015LCID}.

\section{The CARMA merger tree of the Milky Way}\label{sec:merger_tree}

\begin{figure*}[ht!]
\sidecaption
\includegraphics[width=12cm]{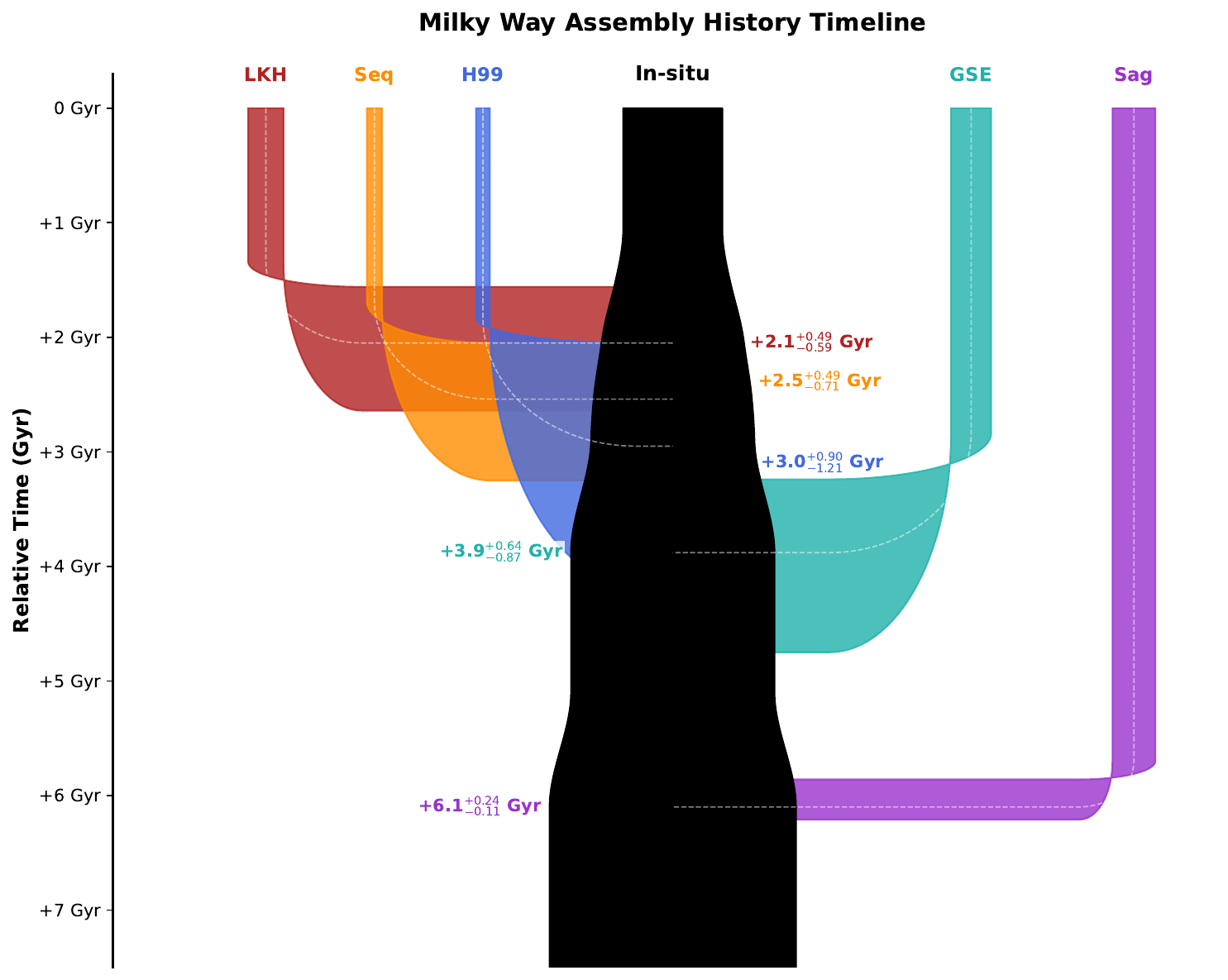}
\caption{Schematic merger tree of the MW revealing the chronological assembly of its main known progenitors (GSE, H99, Seq, and Sagittarius) alongside the LKH population. The main vertical trunk illustrates the in situ formation, with the relative time axis (in Gyr) anchored to $0$~Gyr at $t_i = 14.24$~Gyr as derived in \citetalias{zerbinati26}. For the accreted systems, the width of their incoming vertical stems is scaled proportionally to their total initial stellar mass, $M_\star$, derived from the AMR MCMC modelling of this work, via the yield-mass calibration of \citet[$p \propto M_\star^{0.4}$]{leaman13}, following \citet{massari26}. The transition into horizontal branches marks their integration into the MW, positioned at their MCMC-derived accretion times ($t_f$). The vertical thickness of each branch at the insertion point encodes the corresponding $1\sigma$ temporal uncertainties of the accretion event. For the MW trunk, a mass of $5\times10^{10}\,\mathrm{M}_\odot$ is adopted at the accretion time of GSE \citep{mcmillan17}.}
\label{fig:merger_tree}
\end{figure*}

Figure~\ref{fig:merger_tree} presents the chronological merger tree of the MW as reconstructed from the CARMA AMR modelling of this work. The main vertical trunk represents the main progenitor, with the relative time axis anchored to $t_i = 14.24$~Gyr, which is the onset of in situ GC formation as derived in \citetalias{zerbinati26}. For the MW trunk, a stellar mass of $5\times10^{10}\,\mathrm{M}_\odot$ at the accretion time of GSE is adopted from \citet{mcmillan17}, which corresponds to $p = 0.38$. The four accreted progenitor systems -- GSE, H99, Sagittarius, and Sequoia -- are shown as lateral branches merging into the main trunk at their MCMC-derived accretion times ($t_f$), with branch widths scaled proportionally to their total stellar masses. The vertical thickness of each branch at the insertion point encodes the corresponding $1\sigma$ temporal uncertainties. For consistency, we use $t_f$ as a measure of the accretion time for Sgr, too, even though for a merger so late such an approximation is likely inaccurate (see the discussion in Sect.~\ref{sec:mcmc_results}).  The resulting picture places Sequoia and the Helmi streams as the earliest accreted systems ($\sim11$--$12$~Gyr ago), followed by the dominant GSE merger ($\sim10$~Gyr ago) and the ongoing Sagittarius infall ($\sim8$~Gyr for its last major cluster-forming episode, which might be related to Sagittarius' first pericentric passage), providing the most detailed chronological atlas of the MW's main accretion events to date.

A noteworthy feature of the merger tree is that the accretion events of H99 and GSE are consistent within the uncertainties, suggesting a possible scenario in which H99 was a satellite of the GSE progenitor and was accreted as part of a group-infall event. While dynamical arguments (the prograde H99 orbits vs\ the highly eccentric, radial GSE orbits) favour independent origins \citep{koppelman19}, a definitive answer would require dedicated N-body modelling. The resulting chronological picture places LKH as the earliest merger experienced by the MW, followed soon after (but with significantly smaller contributions in terms of stellar mass) by Sequoia and H99. After a gap of $\sim$1~Gyr, GSE arrives as the dominant merger, followed more than 2~Gyr later by the ongoing Sagittarius infall. This is the most detailed and precise chronological atlas of the MW accretion history constructed to date from strictly homogeneous age information.

We note that our results are in close agreement with the analysis of \citet{lardo26}, who reconstructed progenitor-specific AMRs for 69 MW GCs using a hierarchical Bayesian CMD modelling framework accounting for the presence of multiple stellar populations \citep{valcin26}. Despite this methodological difference, the overall picture emerging from their analysis is qualitatively consistent with ours: their truncation lookback times cluster at $\sim$7--8~Gyr for most systems, Sagittarius achieves the widest metallicity range ($\Delta[\mathrm{Fe/H}]\simeq1.6$~dex), and GSE and the LKH progenitor emerge as the dominant accretion events. The stellar masses they infer for each progenitor match ours remarkably well. Their accretion times broadly overlap with ours, especially given that CARMA ages are on a different absolute scale. When considering the possible systematics at play (see the appendix of \citealt{lardo26}), this qualitative agreement provides independent support for the robustness of the CARMA age-metallicity sequences.

\section{Summary and conclusions}\label{sec:summary}

In this fifth paper of the CARMA series we have extended the homogeneous isochrone-fitting framework of the project to a comprehensive sample of 24 MW GCs dynamically associated with the four main accretion events experienced by the Galaxy outside the LMC and LKH: the GSE system, the Sagittarius dwarf galaxy (Sgr), the Helmi streams (H99), and the Sequoia galaxy (Seq). Relative ages were derived by fitting BaSTI isochrones to deep \textit{HST}/ACS photometry in the F606W and F814W bands, using the CARMA MCMC isochrone-fitting code described in \citetalias{massari23} and \citetalias{aguado25}. The resulting AMRs were then modelled analytically to constrain the mass and accretion time of each progenitor and place them within a coherent merger tree of the MW. The main results of this study can be summarised as follows. Combined with all previous CARMA papers (Papers~I--IV, \citetalias{zerbinati26}, and \citealt{massari26}), this brings the total sample with homogeneous age estimates to 81 GCs, all listed on the CARMA web page\footnote{\url{https://www.oas.inaf.it/en/research/m2-en/carma-en/}}.

\begin{enumerate}

\item \textit{Isochrone-fitting results.} The fitted ages span from $8.07$~Gyr (Terzan~7, Sgr) to $14.16$~Gyr (Terzan~8, Sgr), covering nearly 6~Gyr of SFH across the four progenitor systems. The GSE clusters are the oldest group on average, consistent with the early formation of a massive dwarf galaxy at high redshift. Sagittarius shows the widest internal age spread, from Terzan~8 down to Pal~12 and Terzan~7, reflecting its extended and ongoing enrichment. The fitted photometric parameters ([M/H], E(B$-$V), DM) agree well with the reference values from \citet{harris10} and with the independent distances of \citet{baumgardt21}, confirming the robustness of the CARMA solutions across both photometric catalogues.

\item \textit{AMR-based membership assignments.} To solve the ambiguous dynamical associations of 15 clusters in our sample, we determined a pairwise normalised Mahalanobis likelihood in the age-metallicity plane. The method robustly confirms a GSE membership for 7 clusters ($P(\mathrm{GSE}) \geq0.87$, $N_\sigma>10$) and a H99 membership for 3 clusters (NGC~4590, NGC~5053, and Rup~106). Five GCs have an uncertain AMR association ($N_\sigma<2$) and are assigned to their preferred dynamical progenitor: NGC~3201, NGC~5466, and IC~4499 to Seq, and NGC~6426 and NGC~7078 to H99, bringing the total H99 sample to 5 clusters.

\item \textit{AMR of GSE.} The MCMC model fit to the 11 confirmed GSE members in our sample, supplemented by 11 literature clusters from \citetalias{aguado25} and \citet{massari26}, returns an effective yield of $p = 0.075^{+0.027}_{-0.019}$ and a stellar mass of $M_\star = 8.61^{+9.74}_{-4.36}\times 10^8\,\mathrm{M}_\odot$. Star formation was truncated at $t_f = 10.36^{+0.64}_{-0.87}$~Gyr, consistent within the uncertainties with kinematic and chemical estimates of the GSE merger epoch \citep{helmi18, belokurov18, gallart19}. The inferred stellar mass is consistent with other independent estimates from the literature (\citealt{helmi18, vincenzo19,das20}), confirming the robustness of the method when extended to a broader and more heterogeneous GC sample.

\item \textit{AMR of H99, Sagittarius, and Sequoia.} The three remaining progenitors trace distinct loci in the age-metallicity plane with well-differentiated SFHs. The H99 system is characterised by a low yield ($p = 0.026^{+0.013}_{-0.009}$, stellar mass $M_\star \approx 0.61\times 10^8\,\mathrm{M}_\odot$) and was accreted at $t_f = 11.29^{+0.90}_{-1.21}$~Gyr, about 1.0~Gyr before the GSE event. Sagittarius is the most massive progenitor (stellar mass $M_\star \approx 10.35\times 10^8\,\mathrm{M}_\odot$), with the broadest SFH ($\Delta t \approx 6.7$~Gyr) and the most recent truncation ($t_f = 8.14^{+0.24}_{-0.11}$~Gyr). Sequoia is the lowest-mass progenitor ($p = 0.028^{+0.012}_{-0.008}$, stellar mass $M_\star \approx 0.77\times 10^8\,\mathrm{M}_\odot$), with the shortest SFH ($\Delta t \approx 2.8$~Gyr) and an accretion time of $t_f = 11.70^{+0.49}_{-0.71}$~Gyr.

\end{enumerate}

Taken together, these results provide the first entirely homogeneous and self-consistent chronological picture of the hierarchical assembly of the MW through its main accretion events, as schematised in Fig.~\ref{fig:merger_tree}. The derived accretion timescales place the progenitors in a coherent merger sequence, with LKH, Sequoia, and H99 as the earliest events ($\sim11$--$12$~Gyr ago), followed by the GSE merger ($\sim10$~Gyr ago), and the more extended and recent Sagittarius infall ($\sim8$~Gyr ago for its last major cluster-forming episode). The accretion times and their $1\sigma$ uncertainties are derived directly from the MCMC AMR modelling presented in this work, while the widths of the progenitor branches in the merger tree are scaled proportionally to the total initial stellar masses, $M_\star$, inferred from the same MCMC modelling. We note that the membership assignments presented here represent the most robust approach available to date, as they combine a dynamical preselection in energy--angular-momentum space with a quantitative comparison in the age-metallicity plane. A natural extension of this methodology would be the inclusion of detailed chemical abundances (e.g. [$\alpha$/Fe]) as an additional discriminant, which would further improve the separation of progenitor systems whose AMRs partially overlap. The CARMA project continues to expand its homogeneous GC age library to additional progenitor systems, with the goal of completing a comprehensive chronological atlas of the MW's full accretion history.

\begin{acknowledgements}
We thank the anonymous referee for constructive comments and suggestions that improved the quality of our paper.

D.M.~acknowledges financial support by the Italian Ministry of University and Research (grant FIS2023$-$01611, CUP C53C25000300001) and by INAF through the Mini-Grant n. 1.05.24.07.02.
DM and SC acknowledge financial support from PRIN-MIUR-22: CHRONOS: adjusting the clock(s) to unveil the CHRONO-chemo-dynamical Structure of the Galaxy” (PI: S. Cassisi).
We thank Elena Pancino for her role in establishing the CARMA project and for valuable discussions throughout the series.
CG, TRL and SC acknowledge support from the Agencia Estatal de Investigación del Ministerio de Ciencia e Innovación (AEI-MCINN) under grant "At the forefront of Galactic Archaeology: evolution of the luminous and dark matter components of the Milky Way and Local Group dwarf galaxies in the Gaia era" with reference PID2023-150319NB-C21/C22/10.13039/501100011033. 
TRL acknowledges support from the Ram\'on y Cajal fellowship (RYC2023-043063-I, financed by MCIU/AEI/10.13039/501100011033 and by the FSE+).
YGK acknowledges financial support from PREP2023-001684 funded by MCIU/AEI/10.13039/501100011033 and the FSE+.
SC, DM, MM acknowledge support from the project ``ASTRA: Across Space and Time: Relics \& Archaeology'' (P.I. M. Marconi) funded by INAF-Instituto Nazionale di Astrofisica.
Co-funded by the European Union (ERC-2022-AdG "StarDance: the non-canonical evolution of stars in clusters", Grant Agreement 101093572, PI: E. Pancino and ERC-2024-AdG "ChronoGal: Chronology of our Galaxy from Gaia CMD-fitting", Grant Agreement 101201412). Views and opinions expressed are however those of the author(s) only and do not necessarily reflect those of the European Union or the European Research Council. Neither the European Union nor the granting authority can be held responsible for them.

Based on observations with the NASA/ESA HST, obtained at the Space Telescope Science Institute, which is operated by AURA, Inc., under NASA contract NAS 5-26555. This research made use of emcee \citep{foreman-mackey2013emcee}.
This work has made use of data from the European Space Agency (ESA) mission
\gaia\footnote{\url{https://www.cosmos.esa.int/gaia}}, processed by the \gaia\
Data Processing and Analysis Consortium (DPAC)\footnote{\url{https://www.cosmos.esa.int/web/gaia/dpac/consortium}}. Funding for the DPAC
has been provided by national institutions, in particular the institutions
participating in the \gaia\ Multilateral Agreement.
This project has received funding from the European Research Council (ERC) under the European Union’s Horizon 2020 research and innovation programme (grant agreement No. 804240) for S.S. and \'A.S.
\end{acknowledgements}

\bibliographystyle{aa}
\bibliography{refs}

\begin{appendix}
\onecolumn
\section{MCMC AMR model fits}\label{app:amr_fits}

The MCMC posterior distributions and best-fit AMR curves for each progenitor system are presented below. For each system, the left panel shows the corner plot of the model parameters ($p$, $t_i$, $t_f$ and $M_{\star}$), and the right panel shows the best-fit AMR curve overlaid on the GC data.

\begin{figure}[h!]
    \centering
    \begin{subfigure}{0.42\textwidth}
        \includegraphics[width=\textwidth]{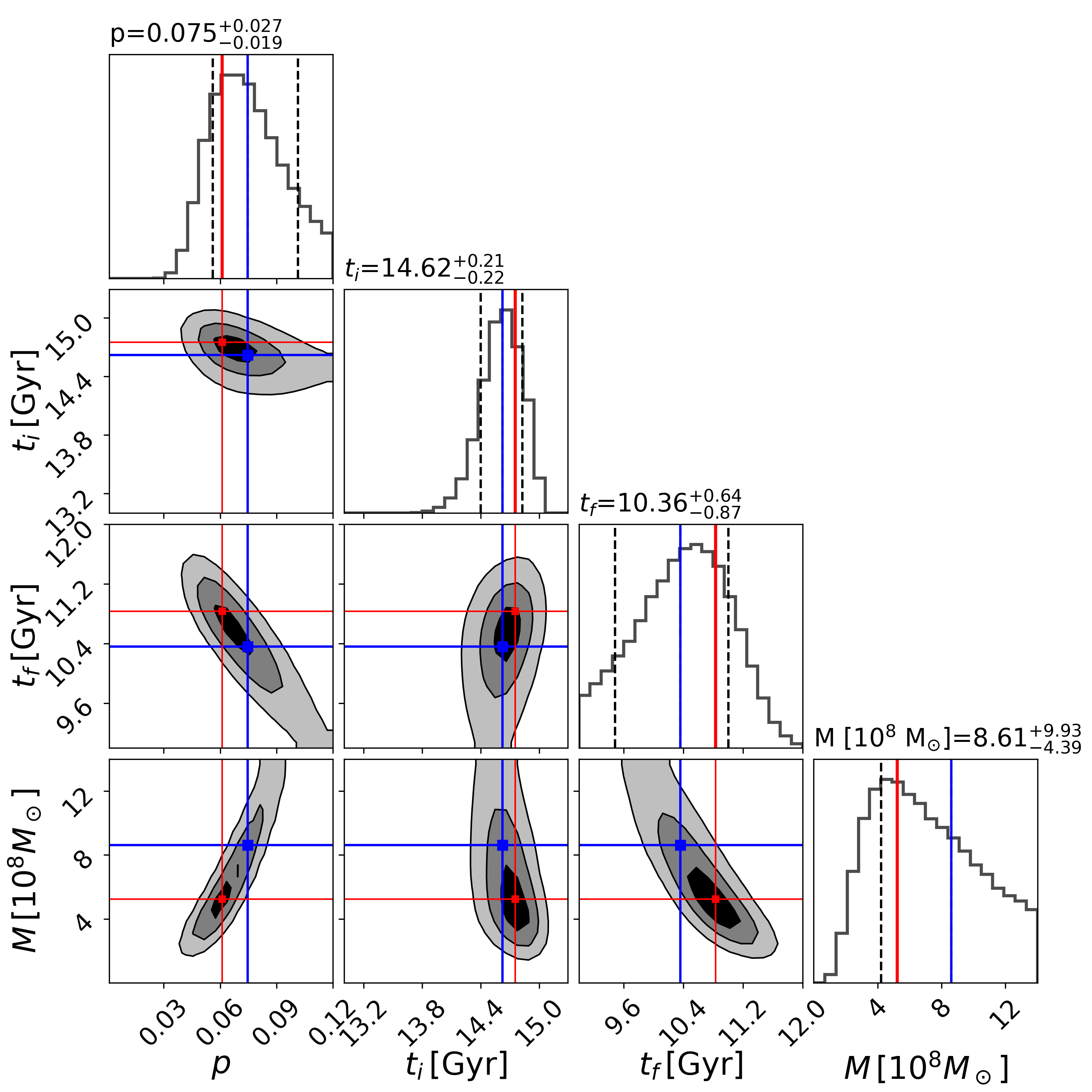}
        
    \end{subfigure}
    \hfill
    \begin{subfigure}{0.42\textwidth}
        \includegraphics[width=\textwidth]{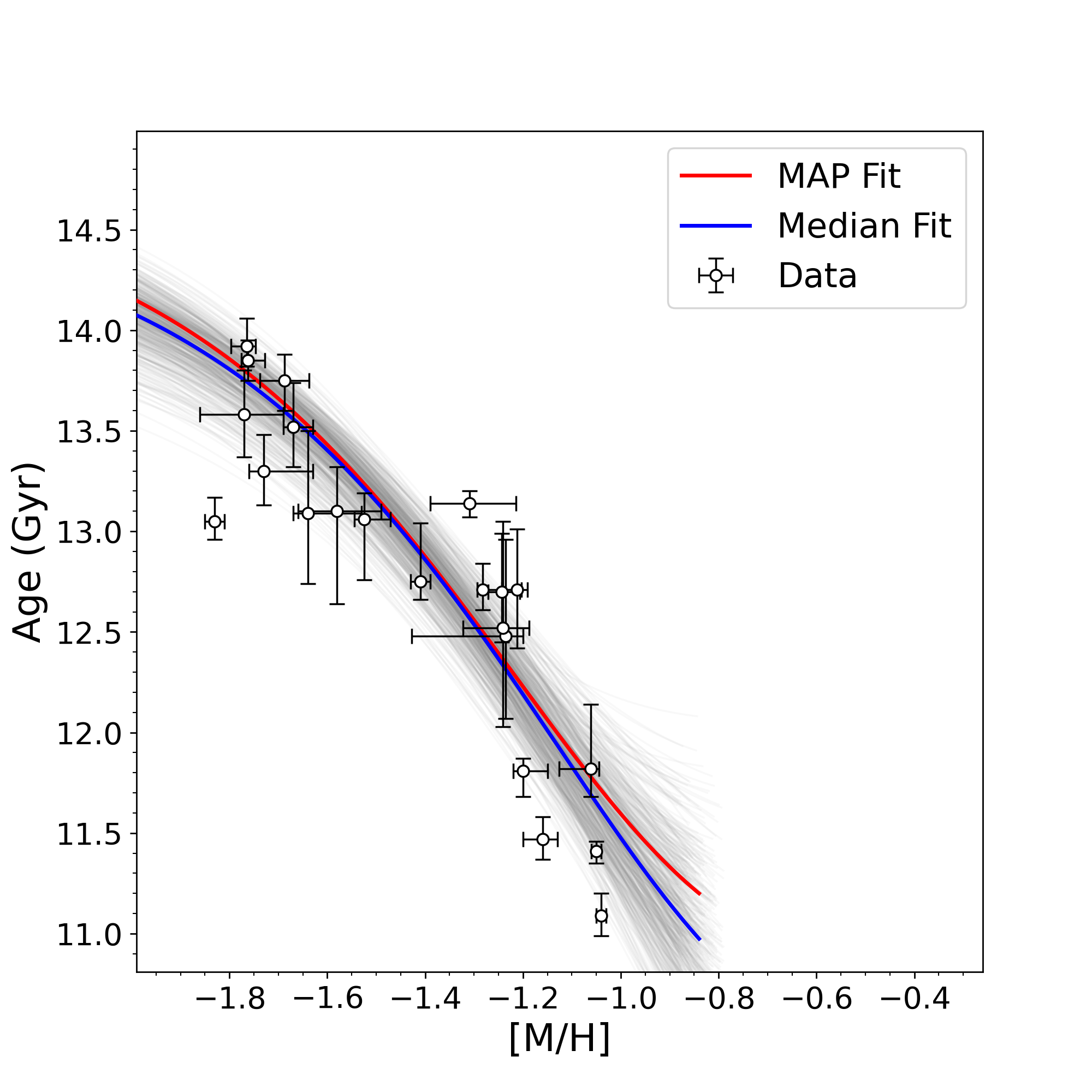}
        
    \end{subfigure}
    
    \vspace{1em}
    \begin{subfigure}{0.42\textwidth}
        \includegraphics[width=\textwidth]{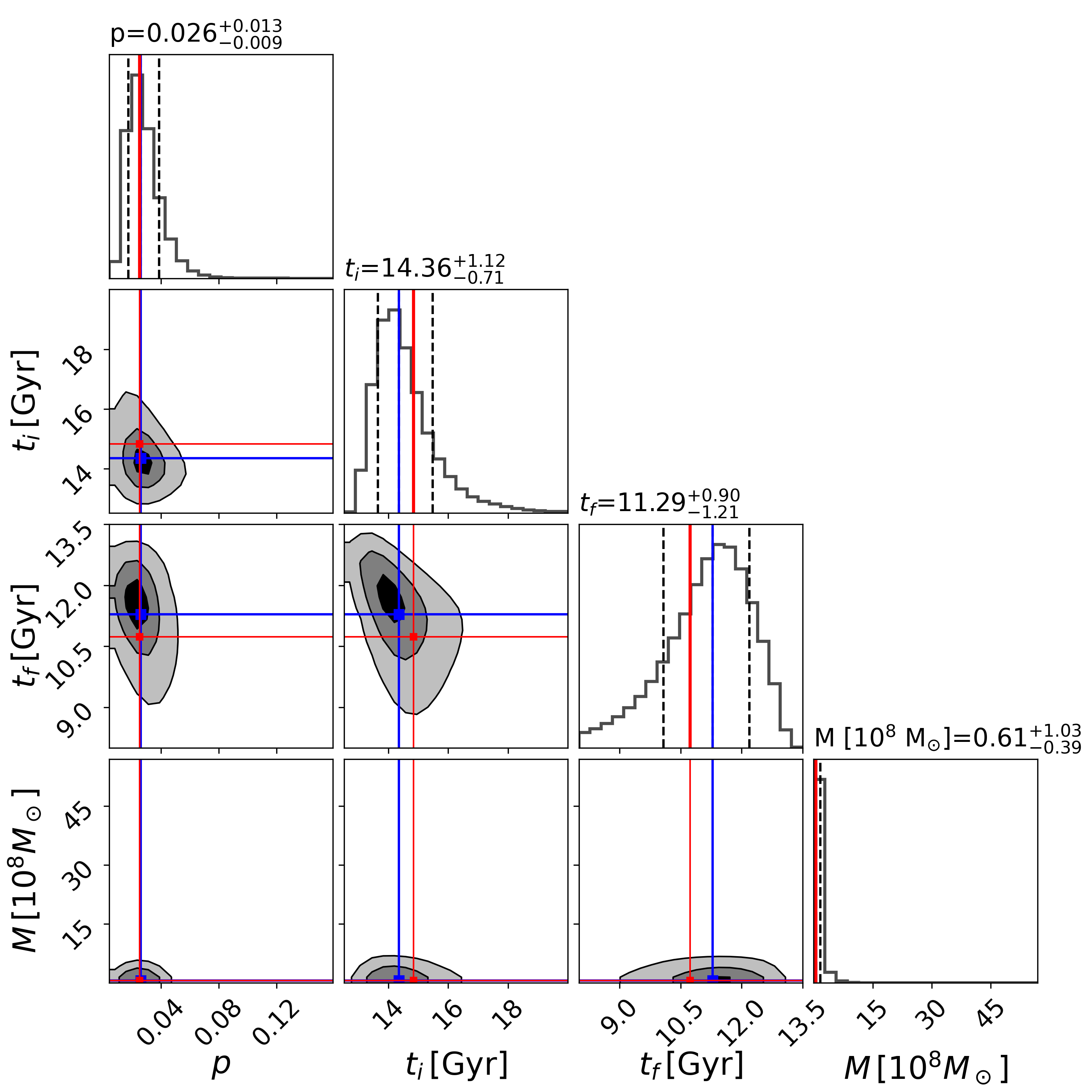}
        
    \end{subfigure}
    \hfill
    \begin{subfigure}{0.42\textwidth}
        \includegraphics[width=\textwidth]{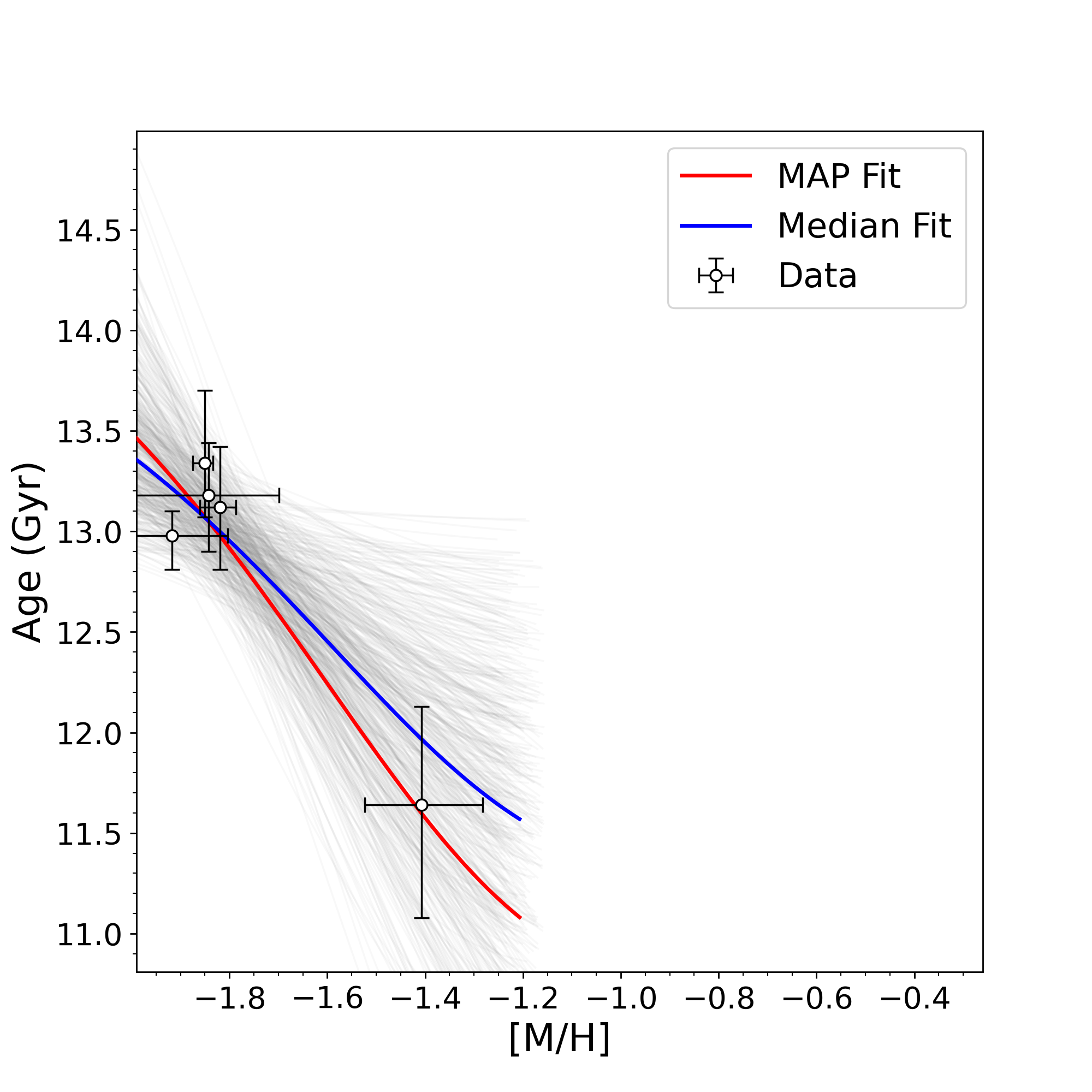}
        
    \end{subfigure}
    \caption{MCMC posterior distributions (\textit{left}) and best-fit AMRs (\textit{right}) for the GSE (\textit{top}) and H99 (\textit{bottom}) progenitor systems.}
    \label{fig:mcmc_amr_models_1}
\end{figure}

\begin{figure}[h!]
    \centering
    \begin{subfigure}{0.42\textwidth}
        \includegraphics[width=\textwidth]{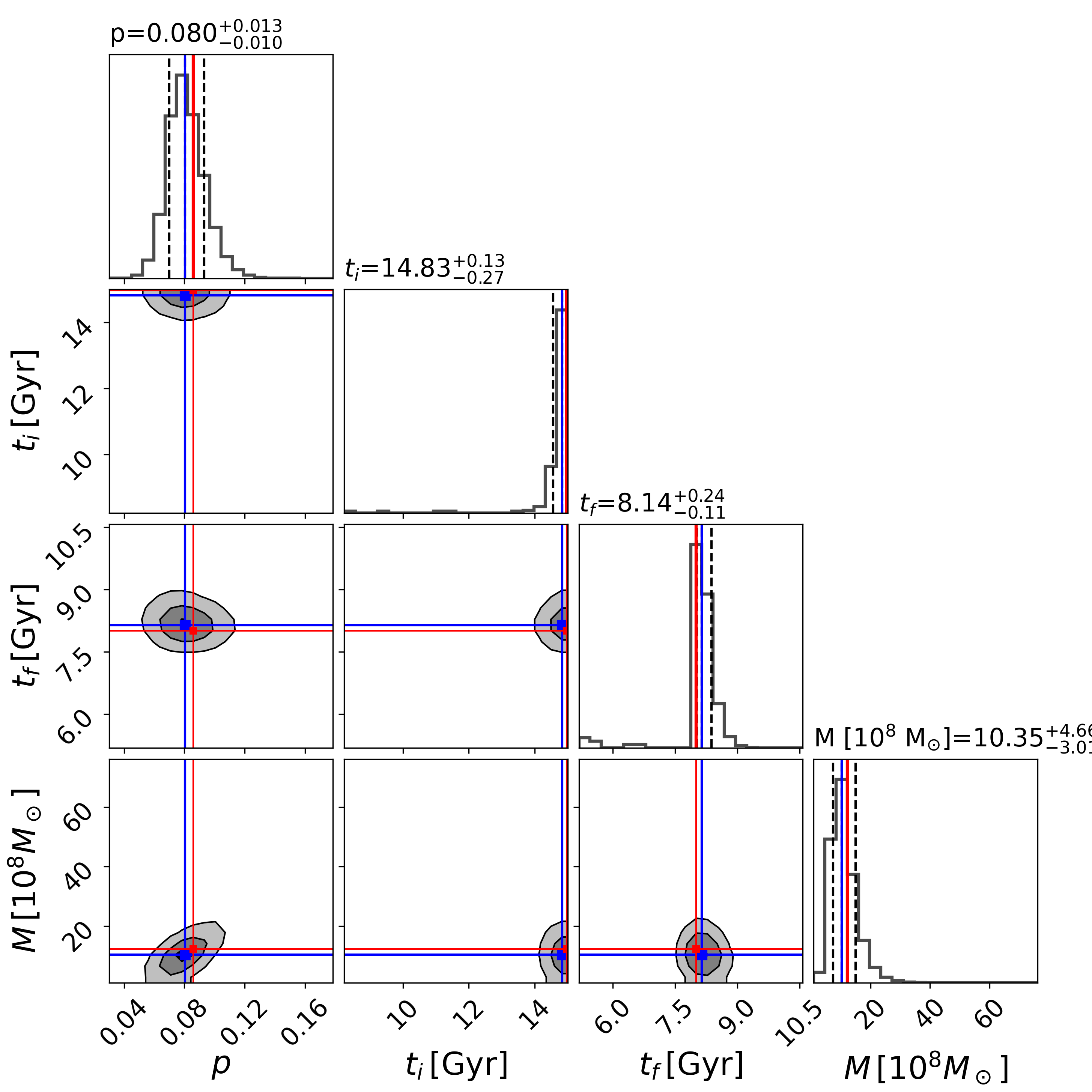}
        
    \end{subfigure}
    \hfill
    \begin{subfigure}{0.42\textwidth}
        \includegraphics[width=\textwidth]{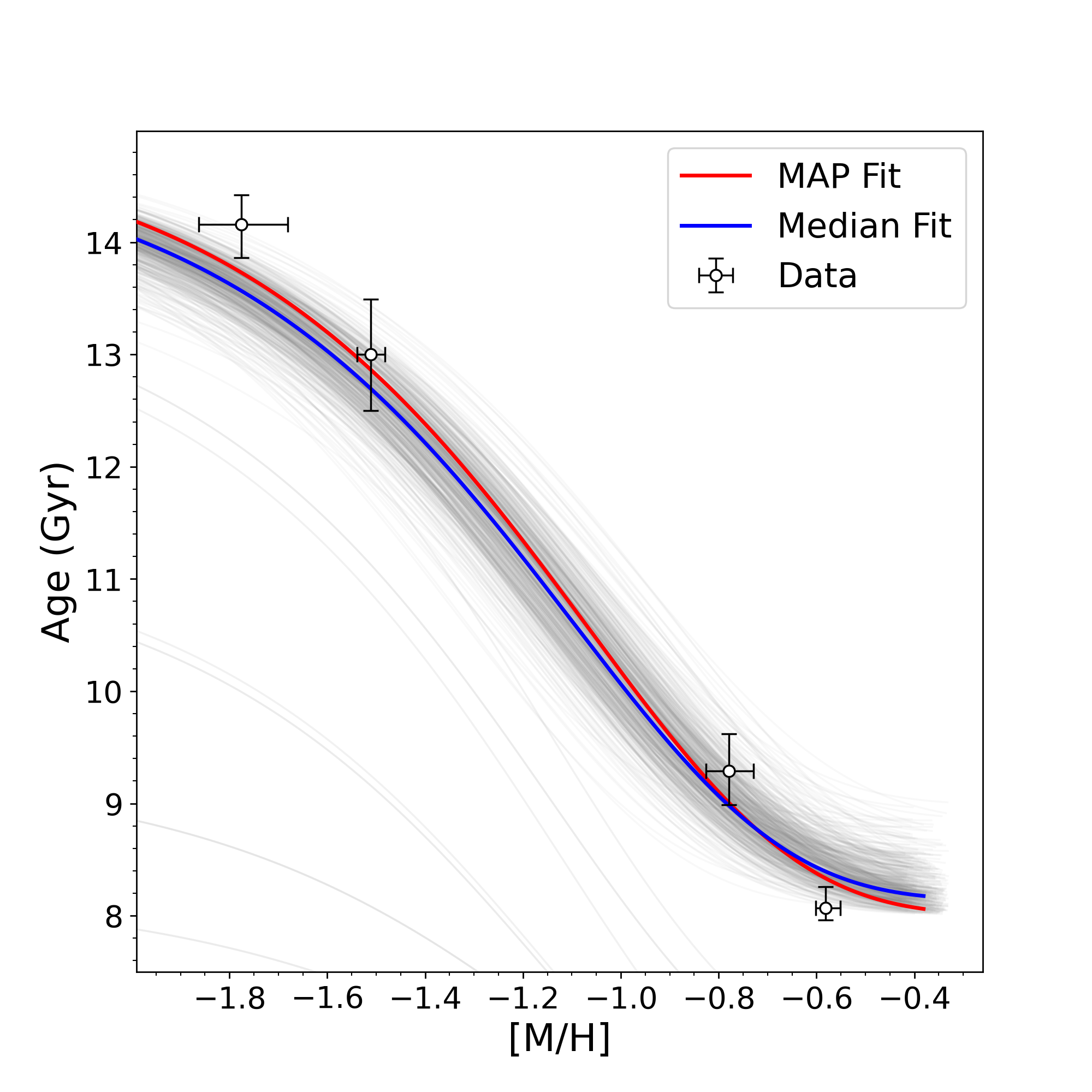}
        
    \end{subfigure}
    
    \vspace{1em}
    \begin{subfigure}{0.42\textwidth}
        \includegraphics[width=\textwidth]{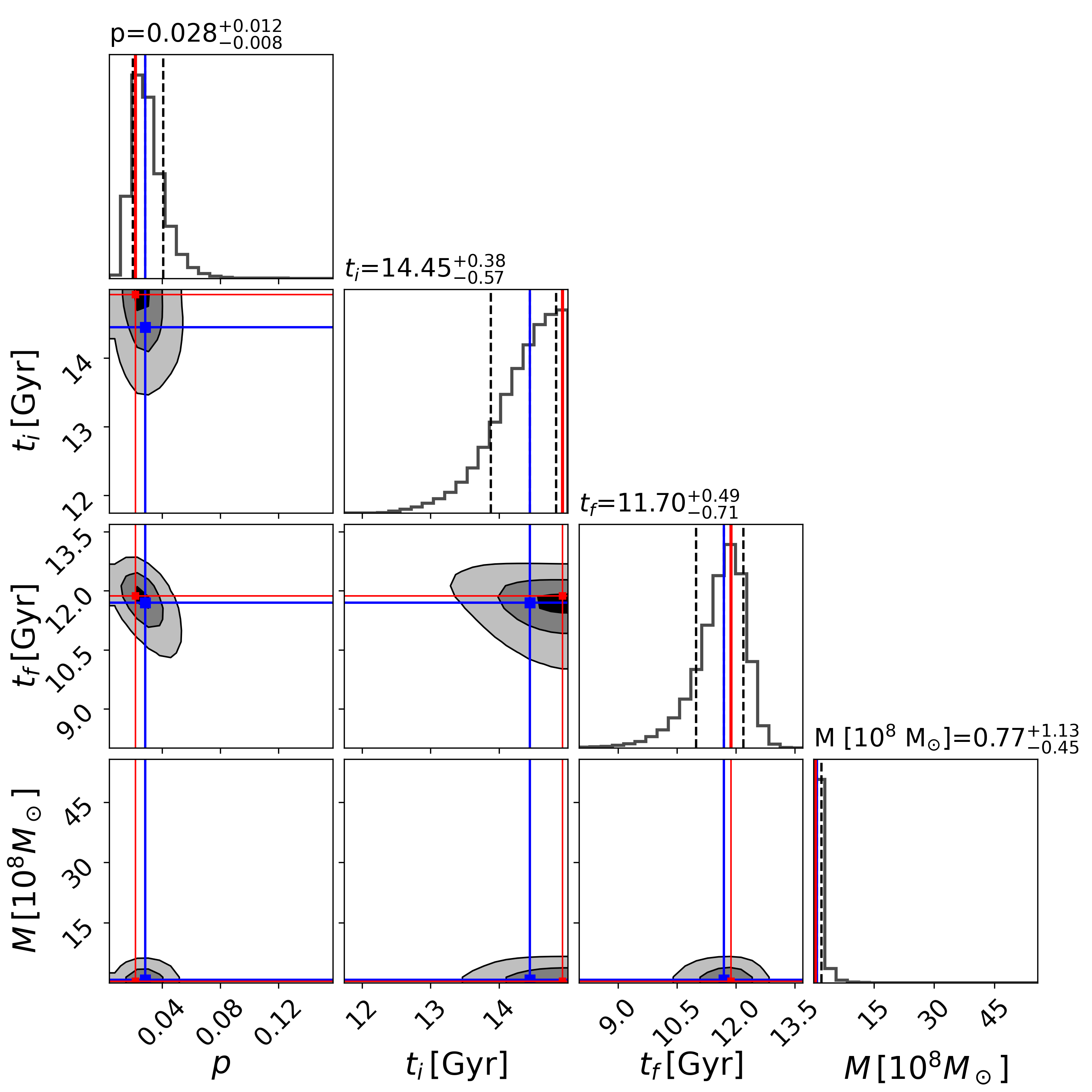}
        
    \end{subfigure}
    \hfill
    \begin{subfigure}{0.42\textwidth}
        \includegraphics[width=\textwidth]{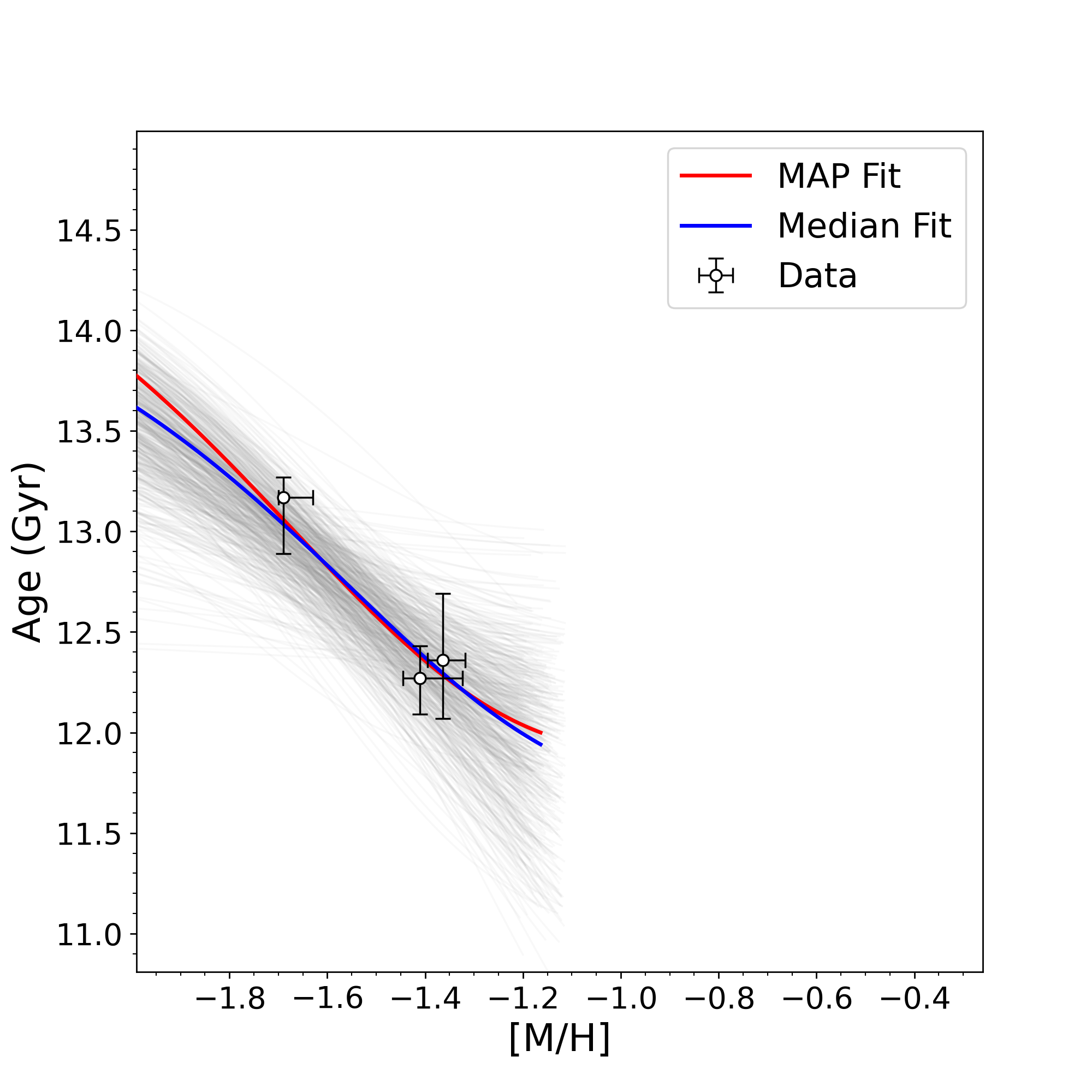}
        
    \end{subfigure}
    \caption{MCMC posterior distributions (\textit{left}) and best-fit AMRs (\textit{right}) for the Sgr (\textit{top}) and Seq (\textit{bottom}) progenitor systems.}
    \label{fig:mcmc_amr_models_2}
\end{figure}

\end{appendix}
\end{document}